\documentclass{ametsoc}
\journal{jcli}

 \usepackage{tabularx}
\usepackage{graphicx}
\usepackage{multirow}
\usepackage{booktabs}
\usepackage[section]{placeins}
\usepackage[export]{adjustbox}
\usepackage{siunitx}
\usepackage{amssymb}
\usepackage{soul}

\DeclareMathOperator*{\argmin}{arg\,min}
\newcommand{\red}[1]{\textcolor{black}{#1}}
\newcommand{\blue}[1]{\textcolor{black}{#1}}
\newcommand{\orange}[1]{\textcolor{black}{#1}}

\usepackage{color}
\usepackage{array}
\newcolumntype{R}[1]{>{\raggedleft\arraybackslash}p{#1}}
\newcolumntype{L}[1]{>{\raggedright\arraybackslash}p{#1}}

\newcommand{\norm}[1]{\left\lVert#1\right\rVert}
\usepackage{enumitem}
\usepackage{float}
\usepackage[utf8]{inputenc}
\DeclareUnicodeCharacter{2061}{}
\title{Regularized Variational Data Assimilation for Bias Treatment using the Wasserstein Metric}
\authors{Sagar K. Tamang\correspondingauthor{Department of Civil, Environmental and Geo-Engineering, University of Minnesota-Twin Cities, Minneapolis, Minnesota}, Ardeshir M. Ebtehaj}
\affiliation{Saint Anthony Falls Laboratory and Department of Civil, Environmental and Geo-Engineering, University of Minnesota-Twin Cities, Minneapolis, Minnesota}
\extraauthor{Dongmian Zou, Gilad Lerman}
\extraaffil{School of Mathematics, University of Minnesota-Twin Cities, Minneapolis, Minnesota}
\email{taman011@umn.edu}

\abstract{
This paper presents a new variational data assimilation (VDA) approach for the formal treatment of bias in both model outputs and observations. This approach relies on the Wasserstein metric stemming from the theory of optimal mass transport to penalize the distance between the probability histograms of the analysis state and an {\it a priori} reference dataset, which is likely to be more uncertain but less biased than both model and observations. Unlike previous bias-aware VDA approaches, the new Wasserstein metric VDA (WM-VDA) dynamically treats systematic biases of unknown magnitude and sign in both model and observations through assimilation of the reference data in the probability domain and can fully recover the probability histogram of the analysis state. The performance of WM-VDA is compared with the classic three-dimensional VDA (3D-Var) scheme on first-order linear dynamics and the chaotic Lorenz attractor. Under positive systematic biases in both model and observations, we consistently demonstrate a significant reduction in the forecast bias and unbiased root mean squared error.}

\begin{document}
\maketitle
\section{Introduction} \label{S:1}
The predictive accuracy of the Earth System Model (ESM) relies on a series of differential equations that are often sensitive to their initial conditions. Even a small error in estimates of their initial conditions can lead to large forecast uncertainties. In short-to-medium range forecast systems, an open-loop run of coupled land and weather models often diverges from the true states and show low forecast skills as the error keeps accumulating over time \citep{charney1951dynamic,kalnay20074}. To extend the forecast lead time, the science of data assimilation (DA) attempts to use the information content of the observations for improved estimates of the ESMs initial conditions thus reducing their forecast uncertainties \citep{leith1993numerical,kalnay2003atmospheric}. The DA methods often involve iterative cycles at which the {\it observations} are optimally integrated with the previous time forecasts ({\it background states}) to obtain an {\it a posteriori} estimate of the initial conditions ({\it analysis state}) with reduced uncertainty in a Bayesian setting \citep{rabier2005overview,asch2016data}. In the literature, two major categories of DA methodologies exist, namely \blue{filtering} and variational methods \citep{law2012evaluating}. Advanced approaches such as hybrid DA schemes are also being developed, which aim to combine and take advantage of unique benefits of the two DA classes \citep{wang2008hybrid,lorenc2015comparison}. 

Although, both classic DA approaches have been widely used in land and weather forecast systems, they are often based on a strict underlying assumption that the error is drawn from a zero-mean Gaussian distribution, which is not always realistic. Bias exists in land-atmosphere models mainly due to under-representation of the governing laws of physics and erroneous boundary conditions. Observation bias also exists largely due to systematic errors in sensing systems and retrieval algorithms represented by the observation operator in DA systems \citep{dee2003detection,dee2005bias}. From a mathematical point of view, bias is simply the expected value of the error that can be removed if the ground truth of the process was known, which is not often feasible in reality. The problem is often exacerbated due to difficulty in attribution of the bias to either model or observations and/or both. \blue{At the same time, point-scale observations such as those from {\it in-situ} gauges and radiosondes are often considered to be more close to the ground truth, however, their assimilation into gridded model outputs is not straightforward due to the existing scale gaps.} 

Bias correction strategies in DA systems mainly fall under two general categories: (i) dynamic bias-aware and (ii) re-scaling bias correction schemes. \orange{Apart from these two general categories, machine learning techniques have also been developed to learn relationships between observations and ancillary variables for bias correction \citep{jin2019machine}. The dynamic bias-aware schemes make prior assumptions about the nature of the bias and attribute it either to model or observations which may not be realistic as both models and observations suffer from systematic errors. Early attempts to dynamically treat model biases are based on a two-step bias estimation and correction approach, which is applied prior to the analysis step \citep{dee1998data,radakovich2001results,chepurin2005forecast}. Variants of bias-aware Kalman filter for colored noise \citep{drecourt2006bias} and a weak constrained four-dimensional VDA (4D-Var) \citep{zupanski1997general} have also been proposed to account for non-zero mean model errors. At the same time, there exists another class of dynamic observational bias correction techniques that rely on variants of variational bias-correction (VarBC) method, which makes an {\it a priori} estimate of bias and dynamically update it using innovation information \citep{auligne2007adaptive,dee2009variational,zhu2014enhanced}. Apart from VarBC, more recently, a new approach is  proposed to treat observation biases by iterative updates of the observation operator \citep{hamilton2019correcting}. A body of research also has been devoted for simultaneously treating model and observation biases using multi-stage hybrid filtering technique \citep{pauwels2013simultaneous}. However, the above schemes still lack the ability to leverage climatologically unbiased information from reference observations (e.g., {\it in situ} data) and has not yet been tested for effective bias correction in chaotic systems. More importantly, the developed schemes largely focus to retrieve an unbiased expected value of the forecast and remain limited to characterization of the second-order forecast uncertainty.}

The re-scaling techniques do not make any explicit assumptions about relative accuracy of the model and observation system \citep{reichle2004bias,crow2005relevance,reichle2007comparison,kumar2009role,reichle2010assimilation,liu2018contributions}. This family of methods often involves mapping the observations onto the model space by matching their Cumulative Distribution Function (CDF). While the CDF-matching technique is comparatively easier in implementation than the dynamic approach and prevents any numerical instabilities in model simulations, it implicitly assumes that model forecasts are unbiased and partly ignores the information content of observations. For example, if our observations were less biased than the model outputs, this approach basically fails to effectively remove the bias. Furthermore, it is a static scheme and there exists no formal way to extend CDF-matching scheme to dynamically account for changes in the bias \citep{kumar2012comparison} and its seasonality \citep{de2016soil}.

Conceptually, CDF-matching techniques move probability masses from one distribution to another. To transform the static CDF-matching to a dynamic scheme, there are two key questions that we aim to answer: Can we quantify the movement of probability masses as a cost through a convex metric? How can this cost be employed to assimilate relatively unbiased {\it in situ} data for dynamic bias correction in the VDA framework? 

The Wasserstein metric (WM) \citep{villani2008optimal,santambrogio2015optimal}, also known as the Earth Mover's distance \citep{rubner2000earth} stems from the theory of optimal mass transport (OMT) \citep{monge1781memoire,kantorovich1942translocation,villani2003topics}, which provides a concrete ground to compare probability measures \citep{brenier1991polar,gangbo1996geometry,benamou2015iterative,chen2017matrix,chen2018optimal,chen2018wasserstein}. Specifically, this distance metric can quantify the dissimilarity between two probability histograms in terms of the amount of ``work'' done during displacement of probability masses between them. Thus, we hypothesize that inclusion of such metric in the VDA cost function can reduce analysis biases. The rationale is that the work done during displacement of the probability masses, is not only a function of the shape of probability histograms but also the difference between their central positions, \blue{as described in Section \ref{sec:OMT}}. The use of the Wasserstein metric in DA has been explored previously \citep{ning2014coping, feyeux2018optimal}. \citet{ning2014coping} introduced the concept of OMT in classical VDA framework and demonstrated that the bias in the background state results in an unrealistic bimodal distribution of the analysis state. However, the study was conducted on linear systems only for model bias correction without accounting for any form of observation biases. \citet{feyeux2018optimal} proposed to fully replace the quadratic costs in the classic VDA by the Wasserstein metric. Even though, the latter approach extends the classic VDA beyond a minimum mean squared error approximation, it does not provide any road map for bias correction, which is the central focus of this paper.

This paper presents a new VDA approach through regularizing the classic VDA problem with cost associated with the Wasserstein metric, hereafter referred to as the Wasserstein Metric VDA (WM-VDA). Unlike previous VDA techniques, WM-VDA treats unknown biases of different magnitudes and signs both in the model dynamics and observations. To that end, WM-VDA needs to be informed by an {\it a priori} reference distribution or histogram (e.g., from {\it in situ} data) that encodes the space-time variability of the state variables of interest in probability domain. This {\it a priori} histogram must be less biased but could exhibit larger higher-order uncertainties than the observations and model forecasts. \blue{More importantly, unlike classic DA methods, the WM-VDA allows full recovery of the probability histogram of the analysis state in the probability domain, which can lead to forecast uncertainty quantification beyond second-order statistics}. The idea is tested on a first-order linear dynamical system as a test-bed and the chaotic Lorenz-63 \citep{lorenz1963deterministic} attractor that represents nonlinear dynamics of convective circulation in a shallow fluid layer. The results demonstrate that the presented approach is capable in preserving the geometric shape of the distribution of analysis state when both the background state and observations are systematically biased and extend the forecast skills by controlling the propagation of bias in the phase space of a highly chaotic system.

The paper is organized as follows: Section 2 discusses the concept of classic VDA focusing on the 3D-Var. In this section, a summary on the theory of OMT and the Wasserstein metric is also provided. The mathematical formulation of the proposed WM-VDA is explained in Section 3. Section 4 implements the WM-VDA on a first-order linear and the nonlinear Lorenz-63 dynamic systems. The results are interpreted and compared with the 3D-Var and \orange{CDF-matching technique}. Summary and concluding remarks are presented in Section 5.

\section{Methodology}
\subsection{Notations}
Throughout, small and capital boldface letters are reserved for representation of $m$-element column vectors $\mathbf{x}\in\mathbb{R}^m$ and $m$-by-$n$ matrices $\mathbf{X}\in\mathbb{R}^{m\times n}$, $\mathbb{1}_m$ is an $m$-element vector of ones and ${\mathbf{I}}_m$ denotes an $m\times m$ identity matrix. A 1-D state variable of interest $\mathbf{x}\in \mathbb{R}$ is represented by a probability vector $\mathbf{p}_x = (p_{x_1},\ldots,p_{x_k})^T$ supported on $k$ points $x_1,\ldots,x_k$, such that $\mathbf{x}=\sum_k p_{x_k} \delta_{x_k}$, where $\delta_{x_k}$ is the Dirac function at $x_k$ and $(\cdot)^T$ denotes the transposition operator. For the state $\mathbf{x}\in \mathbb{R}^m$, this linear expectation operator is represented as $\mathbf{x}=\mathbf{X}\,\mathbf{p}_x$, where the support point and their associated probability of occurrences are properly concatenated in $\mathbf{X}\in \mathbb{R}^{m \times k^m}$ and $\mathbf{p}_x\in\mathbf{R}^{k^m}$, respectively. $\mathbf{x}\sim\mathcal{N}(\boldsymbol{\mu},\,\boldsymbol{\Sigma})$ denotes that $\mathbf{x}$ is drawn from a Gaussian distribution with mean $\boldsymbol{\mu}$ and covariance $\boldsymbol{\Sigma}$, and square of the weighted $\ell_2$-norm of $\mathbf{x}$ is represented as $\norm{\mathbf{x}}_{\mathbf{B}^{-1}}^2=\mathbf{x}^T\mathbf{B}^{-1}\mathbf{x}$, where $\mathbf{B}$ is a positive definite matrix.

\subsection{Classic 3D-Var}

In three-dimensional VDA \citep[3D-Var][]{lorenc1986analysis}, the analysis state is a weighted average of the background state and observations with the weights defined by their respective error covariance matrices. Specifically, let us assume that the $m$-element state variable at time step $t=i+1$ is denoted by $\mathbf{x}_{i+1}\in \mathbb{R}^m$ with the following stochastic dynamics:
\begin{equation}
    \mathbf{x}_{i+1} = \mathbf{\mathcal M}(\mathbf{x}_i)+ \boldsymbol{\omega}_i\,,
\end{equation}
where $\mathbf{\mathcal M} : \mathbb{R}^{m} \xrightarrow{} \mathbb{R}^{m}$ is the nonlinear model operator evolving the state from $\mathbf{x}_i$ to $\mathbf{x}_{i+1}$ and $\boldsymbol{\omega}_i\sim \mathcal{N}(0,\mathbf{B})\in\mathbb{R}^{m}$ is the model error and the expected background state at $i+1$\textsuperscript{th} time step is $\mathbf{x}_b=\mathbf{\mathcal M}( \mathbf{x}_i)$.

Additionally, the $n$-element observation vector available at discrete time step $i$ is denoted by $\mathbf{y}_i \in \mathbb{R}^{n}$ and is related to the true state as follows:
\begin{equation}
    \mathbf{y}_i = \mathbf{\mathcal H}(\mathbf{x}_i) + \boldsymbol{v}_i\,,
\end{equation}
where $\mathbf{\mathcal H} : \mathbb{R}^{m} \xrightarrow{} \mathbb{R}^{n}$ is a nonlinear operator mapping the state space to the observation space and $\boldsymbol{v}_i\sim \mathcal{N}(0,\mathbf{R}) \in \mathbb{R}^n$ is the observation error. 

In a 3D-Var setting, the cost function is comprised of two weighted Euclidean distance of the unknown true state from the background state $\mathbf{x}_b$ and the observation $\mathbf{y}$ such that,
\begin{equation}
\mathcal{J}_{3D}(\mathbf{x}_0) = \norm{\mathbf{x}_0-\mathbf{x}_b}^2_{\mathbf{B}^{-1}}+\norm{\mathbf{y}-\mathbf{\mathcal{H}}(\mathbf{x}_0)}^2_{\mathbf{R}^{-1}}.
    \label{eq:3}
\end{equation}

Assuming that the nonlinear observation operator can be well approximated linearly such that $\mathbf{\mathcal{H}}(\mathbf{x}_i)=\mathbf{H}\,\mathbf{x}_i$, the estimation of the analysis state $\mathbf{x}_a\in \mathbb{R}^m$ at any time step amounts to minimizing the quadratic cost function as follows:
\begin{equation}
\begin{aligned}
    \mathbf{x}_a &=\argmin_{\mathbf{x}_0}\mathcal{J}_{3D}(\mathbf{x}_0)\\ &=\argmin_{\mathbf{x}_0}\bigg\{\norm{\mathbf{x}_0-\mathbf{x}_b}^2_{\mathbf{B}^{-1}}+\norm{\mathbf{y}-{\mathbf{H}}\,\mathbf{x}_0}^2_{\mathbf{R}^{-1}}\bigg\}.
    \end{aligned}
    \label{eq:4}
\end{equation}

Since the minimization problem presented in Equation\,\ref{eq:4} is convex, the local minimum is the global minimum. Thus, by setting the first-order derivative to zero, the analysis state $\mathbf{x}_a$ is obtained as
\begin{equation}
    \mathbf{x}_a =  (\mathbf{H}^T\mathbf{R}^{-1}\mathbf{H}+\mathbf{B}^{-1})^{-1}(\mathbf{H}^{T}\mathbf{R}^{-1}\mathbf{y}+\mathbf{B}^{-1}\mathbf{x}_b),
    \label{eq:5}
\end{equation}

\noindent where the analysis error covariance $\mathbf{P}_a$ is the inverse of the Hessian of Equation\,\ref{eq:3} \citep{daley1993atmospheric}:
\begin{equation}
    \mathbf{P}_a = (\mathbf{H}^{T}\mathbf{R}^{-1}\mathbf{H}+\mathbf{B}^{-1})^{-1}.
\end{equation}

\blue{Clearly the above representations are only valid for a linear observation operator and shall be considered as an an approximate for the nonlinear case. It is also worth mentioning that under the assumption of zero-mean Gaussian errors and linear observation operator, the obtained analysis state, as derived in Equation \ref{eq:5}, can be interpreted as the minimum variance unbiased estimator or the maximum {\it a posteriori} estimator. It is also worth noting that in this setting the results of the 3D-Var is equivalent to the update equations used in standard Kalman filter; however, become suboptimal for non-Gaussian errors and/or a nonlinear observation operator \citep{courtier1994strategy}.}

\subsection{Optimal Mass Transport}
\label{sec:OMT}
The application of the theory of optimal mass transport (OMT) pioneered by Gaspard Monge \citep{monge1781memoire} seems to be a natural extension of the CDF-matching techniques for dynamic bias correction in VDA. The theory was first conceptualized to minimize the total amount of work in transportation of materials between two locations. Recent advances of the theory has provided a fertile ground for comparison of the probability measures \citep{brenier1991polar,villani2003topics} and has been extensively studied and widely applied in the field of signal processing \citep{kolouri2017optimal,motamed2018wasserstein}, image retrieval \citep{rubner2000earth,li2013novel}, and in analyzing misfit in seismic signals \citep{engquist2013application}. 

Let $\mathbf{p}_x=(p_{x_1},\ldots,p_{x_k})^T \in \mathbb{R}^k$ and $\mathbf{p}_{z}=(p_{z_{1}},\ldots,p_{z_{l}})^T \in \mathbb{R}^l$ represent the probability vectors associated with a source and a target histogram supported on vectors $ x_{1},\ldots,x_{k}$ and  $z_{1},\ldots,z_{l}$, respectively. In Monge formulation, the problem involves seeking a surjective optimal transport map $T:\{x_{1},\ldots,x_{k}\}\xrightarrow{}\{z_{1},\ldots,z_{l}\}$ that moves probability mass from each discrete point $x_{i}$ on source probability histogram to a ``single" point $z_{j}$ on target probability histogram, where $i=1\ldots,k$ and $j=1\ldots,l$, such that the total cost of transportation is minimized:
\begin{equation}
    \begin{aligned}
            \underset{T(\cdot)}{\text{minimize}}~\quad &\sum_{i}  c\left(x_{i},T(x_{i})\right)\\
            \textrm{subject to} \quad &{p_{z_{j}}} = \sum_{i: T(x_{i})=z_{j}}{p_{x_i}}\,, 
    \end{aligned}
    \label{eq:7}
\end{equation}
where $c\left(x_{i},T(x_{i})\right)$ is the transportation cost between points $x_{i}$ and $T(x_{i})$ and the constraint warrants the mass conservation principle. Essentially, Monge formulation of OMT problem is non-convex and becomes a combinatorial NP-hard problem, for which the transport map $T(\cdot)$ does not exist -- when target probability histogram has more support points than the source histogram (i.e. $l > k$) \citep{peyre2019computational}. 

\begin{figure}[h]
    \centering
    \includegraphics[width=1\textwidth]{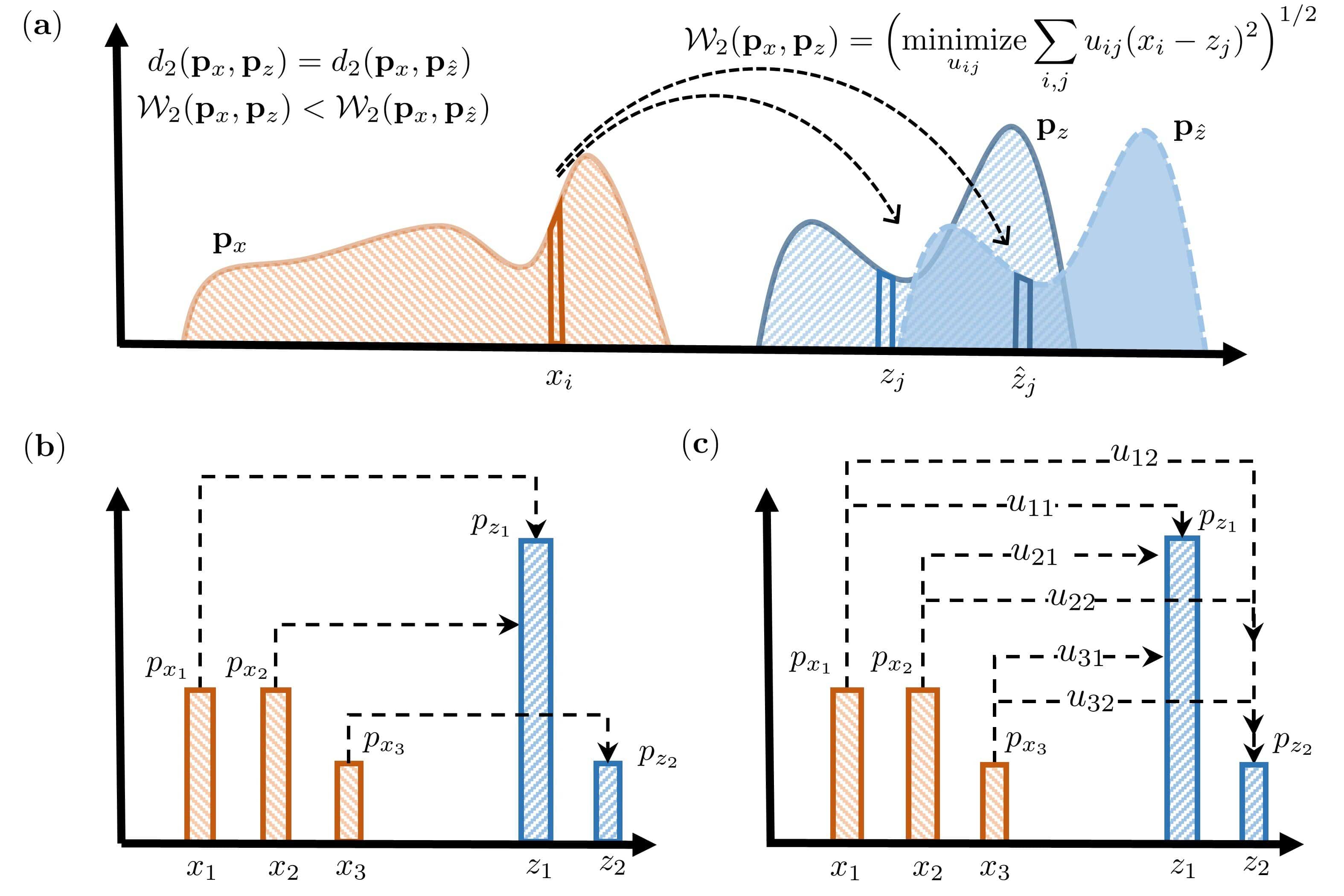}
    \caption{(a) The optimal mass transport problem between source $\mathbf{p}_x$ and two target probability histograms $\mathbf{p}_z$ and $\mathbf{p}_{\hat{z}}$ that are apart from each other only by a first order shift, (b) an example of the Monge formulation of the transportation problem where each source probability mass is assign to only a single target location and (c) the Kantorovich formulation of the problem allowing mass splitting across several target locations. Here, the Euclidean distance ($d_2$) between the source and target probability histograms remains insensitive to the shift in position, while the 2-Wasserstein distance ($\mathcal{W}_2$) increases monotonically as the shift between the central positions of the probability histograms increases.}
    \label{fig:1}
\end{figure}

\citet{kantorovich1942translocation} proposed a convex relaxation of the Monge OMT problem through probabilistic consideration of transport in which mass at any source point $x_{i}$ can be split and transported across several target points $z_{j}$. A schematic of the difference between Monge and Kantorovich formulations of the OMT is shown in Figure\,\ref{fig:1}. Let us assume that $\mathbf{U}\in\mathbb{R}^{k\times l}$ represents the so-called transportation plan matrix, where the element $u_{ij}$ describes the probability mass being transferred from point $x_{i}$ to point $z_{j}$. Here, $\mathbf{C}\in\mathbb{R}^{k\times l}$ represents the transportation or the ground cost matrix, where the element $c_{ij}=|x_{i}-z_{j}|^p$ is the cost of transporting probability masses from $x_{i}$ to $z_{j}$ and $p$ is a positive exponent. Assuming such a ground cost, turns the Kantorovich formulation to the $p$-Wasserstein ($\mathcal{W}_p$) distance or metric, which  seeks to minimize the total amount of work done in transporting probability masses from ${\mathbf p}_x$ to $\mathbf{p}_{z}$ as follows:
\begin{equation}
\centering
    \begin{aligned}
            \mathcal{W}_p(\mathbf{p}_x,\mathbf{p}_{z}) := \Big( \underset{u_{ij}}{\text{minimize}} &\sum_{i,j} c_{ij} u_{ij}\Big)^{1/p} = \Big(\underset{\mathbf{U}}{\text{minimize}}\:\: \textrm{tr}(\mathbf{C}^T\mathbf{U})\Big)^{1/p}\\
                \textrm{subject to}\:\: &\: u_{ij}\ge 0\qquad \\ 
                &\: \mathbf{U}\mathbb{1}_l=\mathbf{p}_x\\
                &\: \mathbf{U}^{T}\mathbb{1}_k=\mathbf{p}_{z}\,.
    \end{aligned}
    \label{eq:8}
\end{equation}

The first constraint assures that the transported probability masses are non-negative and the second and third constraints warrant that the transportation follows conservation of mass. Thus, the Wasserstein metric between two probability histograms is the minimum cost to match them through a transportation plan matrix. Unlike the Euclidean distance that  penalizes the second order statistics of error, the Wasserstein metric penalizes the misfit between the shape of the histograms and increases monotonically with a shift between central position of the histograms \citep{ning2014coping}, enabling it to naturally penalize the bias. \blue{More specifically, for $p=2$, it can be shown that $\mathcal{W}_2^2(\mathbf{p}_x,\mathbf{p}_z) = \mathcal{W}_2^2(\tilde{\mathbf{p}}_x,\tilde{\mathbf{p}}_z) +\norm{\boldsymbol{\mu}_x-\boldsymbol{\mu}_z}^2_2$, where $\tilde{\mathbf{p}}_x$ and $\tilde{\mathbf{p}}_z$ are the centred zero-mean probability masses and $\boldsymbol{\mu}_x$ and $\boldsymbol{\mu}_z$ are the mean values.}

\section{Regularization of VDA through the Wasserstein Metric}

Regularization techniques have been used to reduce uncertainty of the analysis state \citep{wahba1980some,lorenc1986analysis} with isolated singularities \citep{ebtehaj2014variational} ---focusing on precipitation convective cells \citep{ebtehaj2013variational}, sharp transitions in weather fronts \citep{freitag2010l1} and sea ice thickness \citep{asadi2019data}. This paper proposes to add the cost of the square of the $2$-Wasserstein distance as a regularization term to the classic VDA scheme, referred to as WM-VDA. The consideration of $2$-Wasserstein distance also ensures keeping the DA problem convex. This approach not only penalizes the background and observation error in a least-squares sense but also takes into consideration the mismatch between the probability distributions of the analysis state and a ``relatively unbiased'' {\it a priori} reference probability histogram. For example, this histogram can be obtained from the soil moisture gauge measurements, sea ice buoy data or atmospheric radiosondes. Such {\it in situ} measurements are often relatively less biased than both remotely sensed satellite retrievals and the background state, but can exhibit larger higher-order uncertainties, over a sufficiently large window of time or space \citep{villarini2008rainfall}.

\blue{Specifically, let us assume that $\mathbf{p}_{x}= (p_{x_{1}},\dots,p_{x_{k^m}})^T\in \mathbb{R}^{k^m}$ and $\mathbf{p}_{x_r}=(p_{x_{r1}},\ldots,p_{x_{rk^m}})^T\in \mathbb{R}^{k^m}$ represent
the vertically concatenated probability vectors of the analysis state and the {\it a priori} reference data supported on the ``known'' matrix $\mathbf{X}=[{\mathbf x}_{1}|\ldots|\mathbf{x}_{k^m}]\in \mathbb{R}^{m \times k^m}$ where, $k$ represents the number of domain discretization of the state variable in each dimension and $\mathbf{x}_i \in \mathbb{R}^m$ represents the $m$-dimensional vertically concatenated vector of the support points for their associated probability vectors. The WM-VDA cost function is then defined as}
\begin{equation}
    \mathcal{J}_{\text{\tiny WM-VDA}}(\mathbf{x},{\mathbf p}_{x}) = \norm{\mathbf{x}-\mathbf{x}_b}^2_{\mathbf{B}^{-1}}+\norm{\mathbf{y}-{ \mathbf{H}}\mathbf{x}}^2_{\mathbf{R}^{-1}} + \lambda\, \mathcal{J}_{\text{\tiny W2}}\left(\mathbf{p}_{x},\,\mathbf{p}_{x_r}\right)\:,
    \label{eq:9}
\end{equation}

\noindent where $\mathcal{J}_{\tiny \text{w2}}(\mathbf{p}_x,\,\mathbf{p}_{x_r})$ represents the transportation cost associated with the square of the 2-Wasserstein distance between the two probability histograms and $\lambda$ is a non-negative regularization parameter, which balances a trade-off between the Euclidean and the Wasserstein cost. There is no closed form solution to determine this hyper-parameter; therefore, it should be estimated empirically through cross-validation experiments. 

\blue{The analysis state $\mathbf{x}$ at initial time is an expected value that can be represented as $\mathbf{x} = \mathbf{X}\, \mathbf{p}_{x}$. Thus, Equation\,\ref{eq:9} can be expanded as follows:}
\begin{equation}
\blue{    \mathcal{J}_{\text{\tiny WM-VDA}}(\mathbf{p}_{x}) = \norm{ \mathbf{X} \,{\mathbf p}_{x} -{\mathbf x}_b}^2_{\mathbf{B}^{-1}}+ \norm{\mathbf{y}-{\mathbf{H}}\,\mathbf{X} \,{\mathbf p}_{x} }^2_{\mathbf{R}^{-1}} + \lambda\, \mathcal{J}_{\text{\tiny W2}}\left(\mathbf{p}_{x},\,\mathbf{p}_{x_r}\right).}
    \label{eq:10}
\end{equation}

\blue{From the second mass constraint in Equation\,\ref{eq:8}, we have $\mathbf{p}_{x}=\mathbf{U}\mathbb{1}_{k^m}$ and setting $\mathcal{J}_{\text{\tiny W2}}\left(\mathbf{p}_{x},\,\mathbf{p}_{x_r}\right)=\textrm{tr}( \mathbf{C}^T\mathbf{U})$, the above cost function can be expressed in terms of the transportation cost matrix $\mathbf{U}$:}
\begin{equation}
\blue{
    \mathcal{J}_{\text{\tiny WM-VDA}}(\mathbf{U})= \norm{\mathbf{X}\, \mathbf{U}\mathbb{1}_{k^m} -\mathbf{ x}_b}^2_{\mathbf{B}^{-1}}+\norm{\mathbf{y}-{\mathbf{H}}\,\mathbf{X}\, \mathbf{U}\mathbb{1}_{k^m}}^2_{\mathbf{R}^{-1}} + \lambda \:
    \textrm{tr}( \mathbf{C}^T\mathbf{U}).}
    \label{eq:11}
\end{equation}

\blue{Let us assume that $\widetilde{\mathbf{c}}\in \mathbb{R}^{k^{2m}}$ and $\widetilde{\mathbf u}\in \mathbb{R}^{k^{2m}}$ denote lexicographic representation of $\mathbf{C}\in \mathbb{R}^{{k^m} \times {k^m}}$ and $\mathbf{U}\in \mathbb{R}^{{k^m} \times {k^m}}$, respectively, which leads to  $\textrm{tr}(\mathbf{C}^T\mathbf{U})=\widetilde{\mathbf c}^{\,T}\widetilde{\mathbf u}$. Thus, the problem in Equation\,\ref{eq:11} can be recast as a standard quadratic programming problem. To that end, we also need to vectorize the matrix of transportation plan in $\mathbf{p}_{x} = \mathbf{U}\mathbb{1}_{k^m}$ and $\mathbf{p}_{x_r} = \mathbf{U}^T\mathbb{1}_{k^m}$
such that $\mathbf{p}_{x}={\boldsymbol{\Omega}} \widetilde{\mathbf{u}}$ and $\mathbf{p}_{x_r} ={\boldsymbol{\Lambda}} \widetilde{\mathbf{u}}$. Here, $\boldsymbol{\Omega} = [\mathbf{I}_{k^m}|\, \mathbf{I}_{k^m} |\,\ldots |\mathbf{I}_{k^m}] \in \mathbb{R}^{{k^m} \times k^{2m}}$ is the horizontal concatenation of ${k^m}$ identity matrices and $\boldsymbol{\Lambda} = [\mathbf{e}_1|\,\ldots|\mathbf{e}_1| \,\ldots |\mathbf{e}_{k^m}|\ldots|\mathbf{e}_{k^m}]\in \mathbb{R}^{{k^m} \times k^{2m}}$ is the horizontal concatenation of ${k^m}$-dimensional canonical basis, e.g., $\mathbf{e}_1=(1,\ldots,0)^T$ and $\mathbf{e}_{k^m}=(0,\ldots,1)^T$.}\newline
Consequently, Equation\,\ref{eq:11}, can be rearranged as the following standard quadratic programming problem:
\begin{equation}
\blue{    \begin{aligned}
    \mathcal{J}_{\text{\tiny WM-VDA}}(\widetilde{\mathbf{u}})&= \frac{1}{2}\widetilde{\mathbf u}^T\bigg( 2\: \boldsymbol{\Omega}^T\mathbf{X}^T(\mathbf{B}^{-1}+\mathbf{H}^T\mathbf{R}^{-1}\mathbf{H})\mathbf{X}\boldsymbol{\Omega} \bigg)\widetilde{\mathbf u} +
         \bigg(\lambda \: \widetilde{\mathbf c}^T- 2 (\mathbf{x}_b^T\mathbf{ B}^{-1}+\mathbf{y}^T \mathbf{R}^{-1}\mathbf{H}) \mathbf{X}\boldsymbol{\Omega}\bigg)\widetilde{\mathbf u}.
    \end{aligned}}
    \label{eq:12}
\end{equation}
Thus, the WM-VDA amounts to obtaining the ``analysis transportation plan'' $\widetilde{\mathbf{u}}_a$:
\begin{equation}
    \begin{aligned}
        \widetilde{\mathbf{u}}_a =\quad& \underset{\widetilde{\mathbf{u}}}{\text{argmin}}
        & & \mathcal{J}_{\text{\tiny WM-VDA}}(\widetilde{\mathbf{u}}) \\
        & \text{subject to}
        & & \widetilde{u}_{i}\ge 0\\
        &&& {\boldsymbol{\Lambda}} \widetilde{\mathbf{u}} = \mathbf{p}_{x_r},
    \end{aligned}
    \label{eq:13}
\end{equation}
\noindent  which can be efficiently solved through interior-point optimization techniques \citep{altman1999regularized}. \blue{Finally, the analysis state can be obtained as $\mathbf{x}_a = \mathbf{X}\,{\boldsymbol{\Omega}}\, \widetilde{\mathbf{u}}_a$. It shall be noted that, since the obtained analysis state doesn't exactly satisfy the constraint of Wasserstein metric, WM-VDA is a weak-constraint DA formulation based on the terminology introduced by \citet{daley1993atmospheric}.}

\section{Numerical Experiments and Results}

In DA experimentation, we run the forward model under controlled model and observation error, which enables to characterize the effectiveness of the proposed methodology in comparison with the classic 3D-Var approach. To initially examine the performance of the WM-VDA, we focus on two dynamic systems with different levels of sensitivity to their initial conditions including a first order linear and the chaotic Lorenz-63 system. First order dynamical systems have been the cornerstone in developing the Kalman filter \citep{kalman1960new} and has been widely used as a test-bed to examine the performance of new filtering techniques \citep{hazan2017learning,hazan2018spectral}. On the other hand, the Lorenz-63 has been the subject of numerous experiments to test the performance of new DA techniques under a chaotic dynamics \citep{anderson1999monte,miller1994advanced,harlim2007non,van2010nonlinear,reich2012gaussian,goodliff2015comparing}.

\subsection{First Order Linear Dynamics}
\subsubsection{State-space characterization}

A first-order discrete time representation of a linear dynamic system in the state-space is presented as follows:
\begin{align}
\mathbf{x}_{i+1}= \,& \mathbf{M}\,\mathbf{x}_{i}+\boldsymbol{w}_{i}\nonumber\\
\mathbf{y}_{i}= \,& \mathbf{H}\,\mathbf{x}_{i}+\boldsymbol{v}_{i}\,, 
\label{eq:14}
\end{align}

\noindent where, $\mathbf{M}\in\mathbb{R}^{m\times m}$ is the (time-invariant) state transition matrix. It is important to note that the system remains stationary if and only if $\underset{i}{\max}\left\{|\gamma_i|\right\}<1$, where $\left\{\gamma_i\right\}_{i=1}^m$ denote the eigenvalues of $\mathbf{M}$. To examine the effectiveness of the proposed WM-VDA approach in treating systematic biases, we introduce a shift in the system dynamics by assuming that the model and observation errors are drawn from non-zero mean Gaussian distributions.

\subsubsection{Assimilation set-up and results}
Here, we confine our considerations to a 1-D simulation of a linear state space. The initial parameter values were chosen as $\mathbf{x}_0=10$ and $\mathbf{M}=0.97$. The expected value (ground truth) of the true state trajectory is generated by solving the model dynamics in Equation\,\ref{eq:14} with a time step of $\Delta t=0.01$ [t] over a period of $T=3$ [t], in the absence of any model error. For each simulation time step, the model error is drawn from $\boldsymbol{\omega}_i\sim\mathcal{N}(0.5,\,1.5)$. The observations are obtained by corrupting the truth at assimilation interval $T_a= 3\Delta t$ using $\mathbf{v}_{i}\sim\mathcal{N}(0.25,\,0.75)$. To represent the relatively unbiased but highly uncertain {\it a priori} reference probability histogram $\mathbf{p}_{x_r}$, 500 samples are drawn at each assimilation intervals from a Gaussian distribution where its mean is located on the ground truth and its variance is set to $\boldsymbol{\sigma}^2_{x_r}=3\sigma^2_{b}$, where $\sigma^2_{b}=1.5$ is the background error covariance. To solve the WM-VDA, the regularization parameter is set to $\lambda =5$. As will be explained later, this value empirically minimizes the analysis mean squared error (MSE). In order to have a robust conclusion about comparison of the proposed WM-VDA scheme with the classic 3D-Var, the DA experimentation is repeated for 50 independent ensemble simulations. 

\begin{figure}[h]
    \includegraphics[width=1\textwidth]{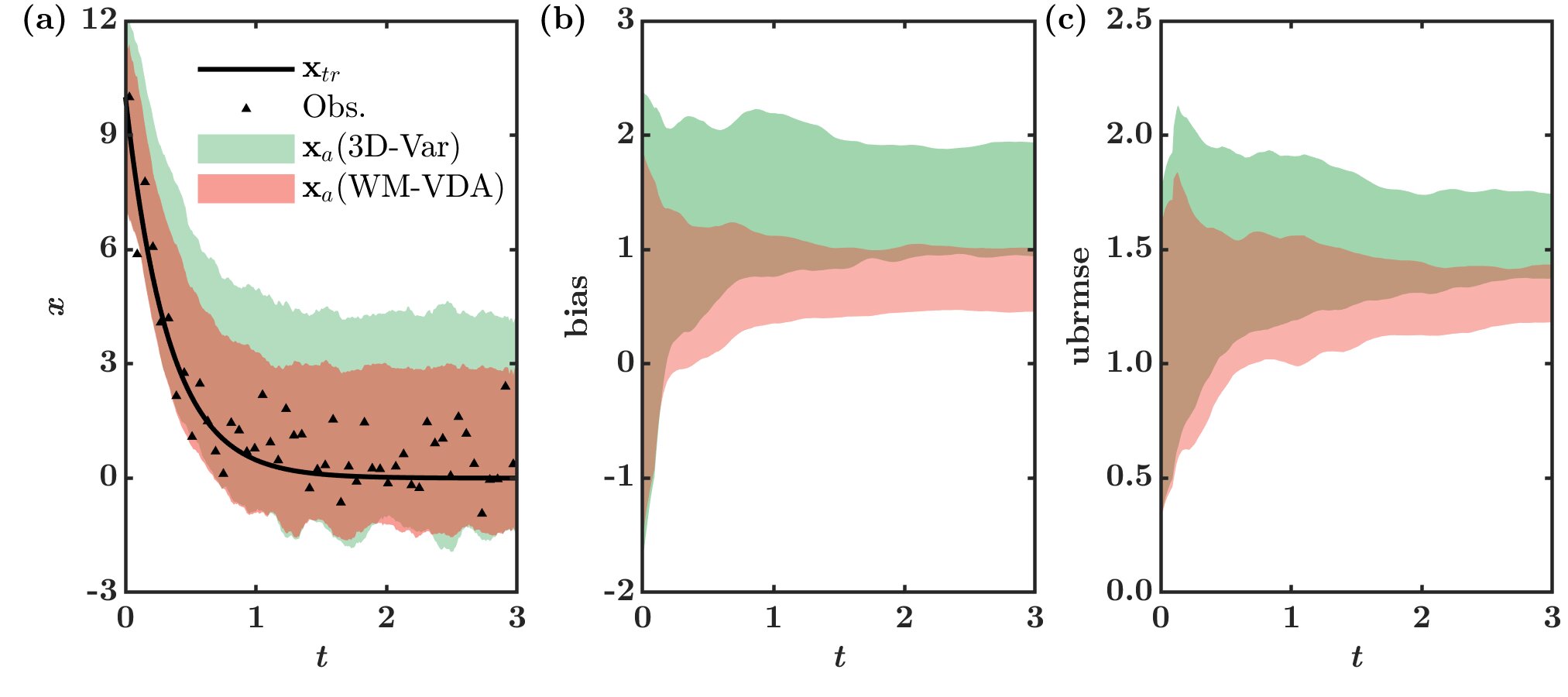}
    \caption{Time evolution of the uncertainty range (colored shaded region) for the 3D-Var (green) and WM-VDA (orange)---representing (a) the 2.5 and 97.5 percentiles of 50 ensemble members of the analysis state $\mathbf{x}_a$ and the 95\% confidence bound for its (b) bias and (c) ubrmse. The true state ($\mathbf{x}_{tr}$) and observations (Obs.) for an independent simulation are shown in the left panel as well.}
    \label{fig:2}
\end{figure}

For all 50 ensembles, the ground truth of the model trajectory, the 2.5 and 97.5 percentiles of the ensemble members and their associated quality metrics, including the bias and unbiased root mean squared error (ubrmse), are shown in Figure\,\ref{fig:2}. As is evident, not only the uncertainty range of the ensemble members (Figure\,\ref{fig:2}\,a) but also the quality metrics (Figure\,\ref{fig:2}\,b and c) noticeably improved for the WM-VDA scheme compared to the 3D-Var. In particular, on average, WM-VDA leads to the reduction of bias (ubrmse) from 1.4 to 0.7 (1.6 to 1.3), which is equivalent to 50\% (19\%) reduction when compared with the 3D-Var.

\begin{figure}[h]
    \includegraphics[width=1\textwidth]{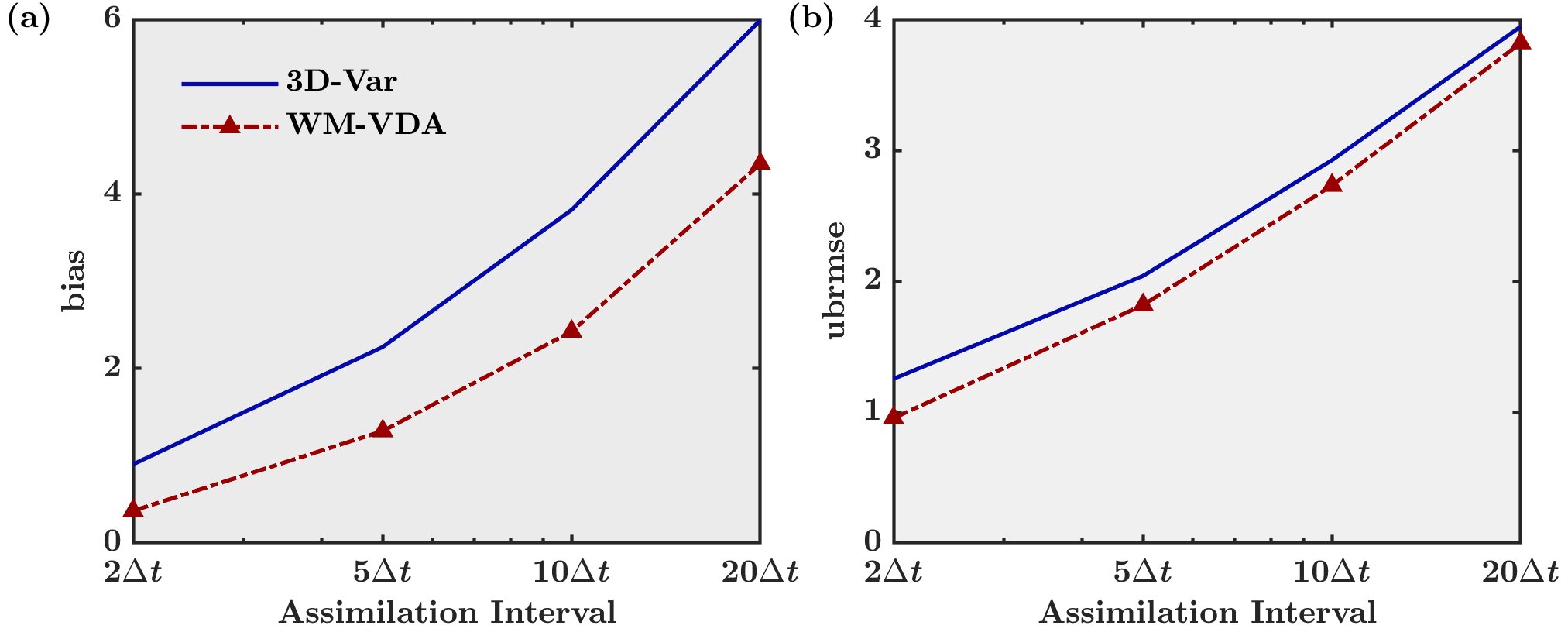}
    \caption{Variation of (a) bias and (b) ubrmse for the 3D-Var and WM-VDA as a function of assimilation interval.}
    \label{fig:3}
\end{figure}

The sensitivity of the quality metrics is also tested for different range of assimilation intervals $T_a = \{2\Delta t,\,5\Delta t,\,10\Delta t,\,20\Delta t\}$ for both DA schemes (Figure\,\ref{fig:3}\,a and b). It is found that the WM-VDA improves the error quality metrics compared to 3D-Var across the chosen range of the assimilation intervals. In particular, for small assimilation intervals of 2$\Delta t$, WM-VDA reduces the bias (ubrmse) from 1 to 0.5 (1.25 to 1), which is equivalent to 50\% (20\%) reduction when compared to classic 3D-Var. As expected, the ubrmse is reduced less significantly than the bias as the variance of the assimilated reference probability histogram was markedly larger than both observations and background state. As shown, when the assimilation interval grows, the bias and ubrmse in both schemes monotonically increase; however, at different rates. In fact, the bias grows faster in 3D-Var than the WM-VDA while the already small gap between the ubrmse values from the two methods slightly shrinks as the assimilation interval grows. Thus, the analysis state bias in WM-VDA seems to be more robust to increased assimilation intervals than the 3D-Var. This feature needs further investigation as it could be highly desirable for land surface DA, since the satellite overpasses are often available at much longer time intervals than the forecast time steps. 

As previously noted, the regularization parameter $\lambda$ plays a significant role in WM-VDA algorithm by making a trade-off between the weighted Euclidean cost and the transportation cost. Recall that larger values of $\lambda$ push \blue{or overfits} the analysis state towards the {\it a priori} reference probability histogram by neglecting the information content of background state and observations and thus reduce bias at the expense of an increased spread in uncertainty of the analysis state. On the other hand, smaller values diminish the role of the transportation cost and render it ineffective for bias correction. As previously noted, there is no closed form solution for optimal approximation of this parameter. Here, we focus on determining optimal values for $\lambda$ through cross-validation and trial and error analysis. 

\begin{figure}[h]
    \includegraphics[width=1\textwidth]{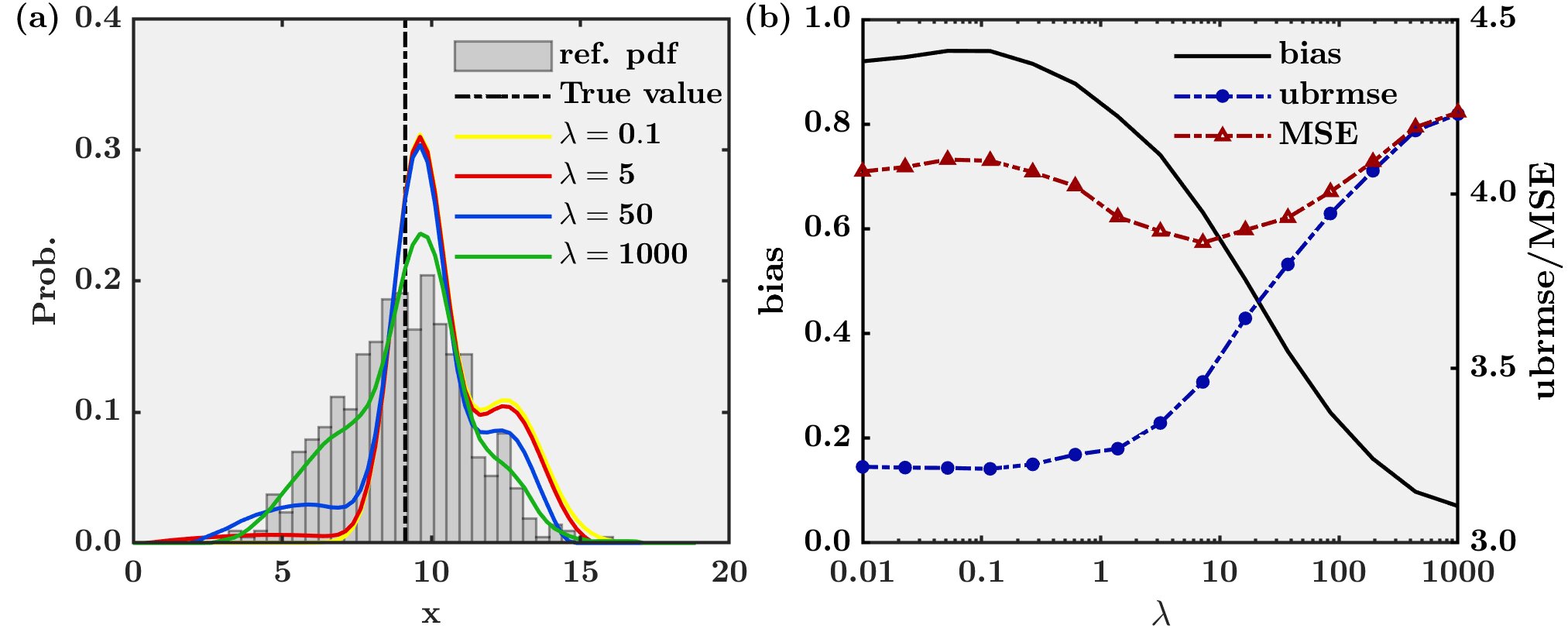}
    \caption{(a) Transition of the analysis state probability histogram at first assimilation cycle and (b) variation of bias, ubrmse and mean squared error (MSE) of the analysis error as a function of $\lambda$, averaged over 50 independent ensemble simulations.}
    \label{fig:4}
\end{figure}

Figure\,\ref{fig:4}\,a demonstrates the evolution of the analysis probability histogram as a function of the regularization parameter $\lambda = \{0.1, 5, 50, 1000\}$, at the first assimilation cycle, for the shown experimental setting in Figure\,\ref{fig:2}. It can be seen that for small values of $\lambda \le 5$, due to the existing biases, the analysis probability histogram acquires a bimodal distribution as the results approach to those of the 3D-Var scheme. However, as the value of $\lambda$ is increased, the bimodality begins to fade and a more proper geometry of the analysis probability histogram is recovered. At larger values of $\lambda$ = 1000, the analysis probability histogram matches perfectly with the {\it a priori} reference probability histogram as the transportation cost dominates the quadratic cost of the 3D-Var.

Figure\,\ref{fig:4}\,b shows the variation of the quality metrics as a function of $\lambda$ averaged over 50 independent ensemble simulations. It can be seen that the bias decreases and ubrmse increases for larger values of $\lambda$, yielding a trade-off point where the MSE is minimal. As shown, this minimum MSE is achieved in the range of $\lambda =$ 5--10 based on the chosen model parameters and error terms. Clearly, for every problem at hand, this analysis needs to be done offline prior to the implementation of the WM-VDA scheme. 

\subsection{Lorenz-63} 
\subsubsection{State-space characterization}

Lorenz system \citep[Lorenz-63,][]{lorenz1963deterministic} is a chaotic ordinary differential equation obtained as Fourier truncation of the Rayleigh-B\'{e}nard convective flow of fluids in which the three coordinates $x$, $y$ and $z$ represent rate of convective overturn, horizontal and vertical temperature variations, respectively. The system is represented as follows:

\begin{equation}
    \begin{split}
        \frac{dx}{dt} &= -\sigma (x-y)\\
        \frac{dy}{dt} &= \rho x-y-xz\:\:,\\
        \frac{dz}{dt} &= xy-\beta z\\
    \end{split}
    \label{eq:15}
\end{equation}

where, $\sigma$, $\rho$ and $\beta$ are the Prandtl number, a normalized Rayleigh number and a dimensionless wave number respectively. The standard parameter values are $\sigma = 10$, $\rho=28$ and $\beta = 8/3$ for which the system exhibits a strong chaotic behaviour with maximum Lyapunov exponent value of 0.9. For this selection of parameters, there exist three equilibrium points: the origin that represents the conductive state of no motion and the two attractors located at ($\sqrt{\beta(\rho-1)},\sqrt{\beta(\rho-1)},\rho-1$) and ($-\sqrt{\beta(\rho-1)},-\sqrt{\beta(\rho-1)},\rho-1$),  representing a pattern of convective rolls with different directions of rotation.


\subsubsection{Assimilation setup and results}
In this subsection, we present the results from two different DA settings for the Lorenz system, which differ in terms of characterization of the {\it a priori} reference probability histogram $p_{x_r}$. In setup I, it is assumed that $p_{x_r}$ is available at each assimilation interval while in setup II, it is available over a window of time. The reason for examination of the second experimental setup is that, in practice, adequate samples to construct the reference histogram may not be available at each assimilation cycle. However, histogram of historical {\it in situ} observations is often available over a window of time or space that can be leveraged to reduce the bias in DA systems. An example is the availability of monthly or seasonal probability histogram of gauge measurements of surface soil temperature, moisture or radiosonde data of atmospheric states. \blue{To minimize the computational cost of interior-point optimization algorithm, we solved the problem of determining analysis state in each dimension separately by setting $m=1$ and utilizing marginal probability histograms along each dimension.}

In both experiments, to obtain the ground truth of the model trajectory, the Lorenz system is initialized at ($x$, $y$, $z$) = (3,\,$-$3,\,12) and is integrated using a fourth-order Runge-Kutta \citep[RK4,][]{runge1895numerische,kutta1901beitrag} approximation with a time step of $\Delta t= 0.01$ [t] over a time period of $T = 0-50$ [t] at the end of which the system attains second order stationarity. The observations are obtained at every $10 \Delta t$ by corrupting the truth over a simulation period of $T = 0-20$ [t] using a Gaussian error $v_i \sim \mathcal{N}(\boldsymbol{\beta}_y\mathbb{1}_3,\,\sigma_y^2\, \mathbf{I}_3)$ where $\boldsymbol{\beta}_y=0.15$ and $\sigma_y^2=2$. The model is propagated upto $T =20$ [t] by adding a Gaussian noise $\boldsymbol{\omega}_i \sim \mathcal{N}(\boldsymbol{\beta}_m\mathbb{1}_3,\,\sigma_b^2\, \mathbf{I}_3)$ to the model state at every $\Delta t$, where $\boldsymbol{\beta}_m=0.25$ and $\sigma_b^2= \sqrt 5$. For setup I, at the assimilation interval of $T_a=10\Delta t$, 500 samples are drawn from a Gaussian distribution with mean of the ground truth and covariance $\sqrt 3\,\sigma_b^2 \mathbf{I}_3$ to construct the unbiased {\it a priori} reference probability histogram $p_{x_r}$. To solve the WM-VDA, the regularization parameter is set to $\lambda = 3$ for each dimension by trial and error. However, in setup II, $p_{x_r}$ is constructed through adding a zero-mean Gaussian noise with covariance $\sqrt3\,\sigma_{b}^2 \mathbf{I}_3$ to the ground truth over a period of $T = 0-50$ [t]. In this case, the $p_{x_r}$ has a larger spread than the setup I and provides an unbiased representation of the process over the entire simulation period. In setup II, the regularization parameter is set to $\lambda = (0.02,\,0.08,\,0.07)$ for the three coordinates. Throughout, in order to have a robust conclusion about comparison of the proposed WM-VDA scheme with the 3D-Var, DA experiments are repeated for 50 independent ensemble simulations for both experimental settings.

\begin{figure}[h]
    \centering
    \includegraphics[width=1\textwidth]{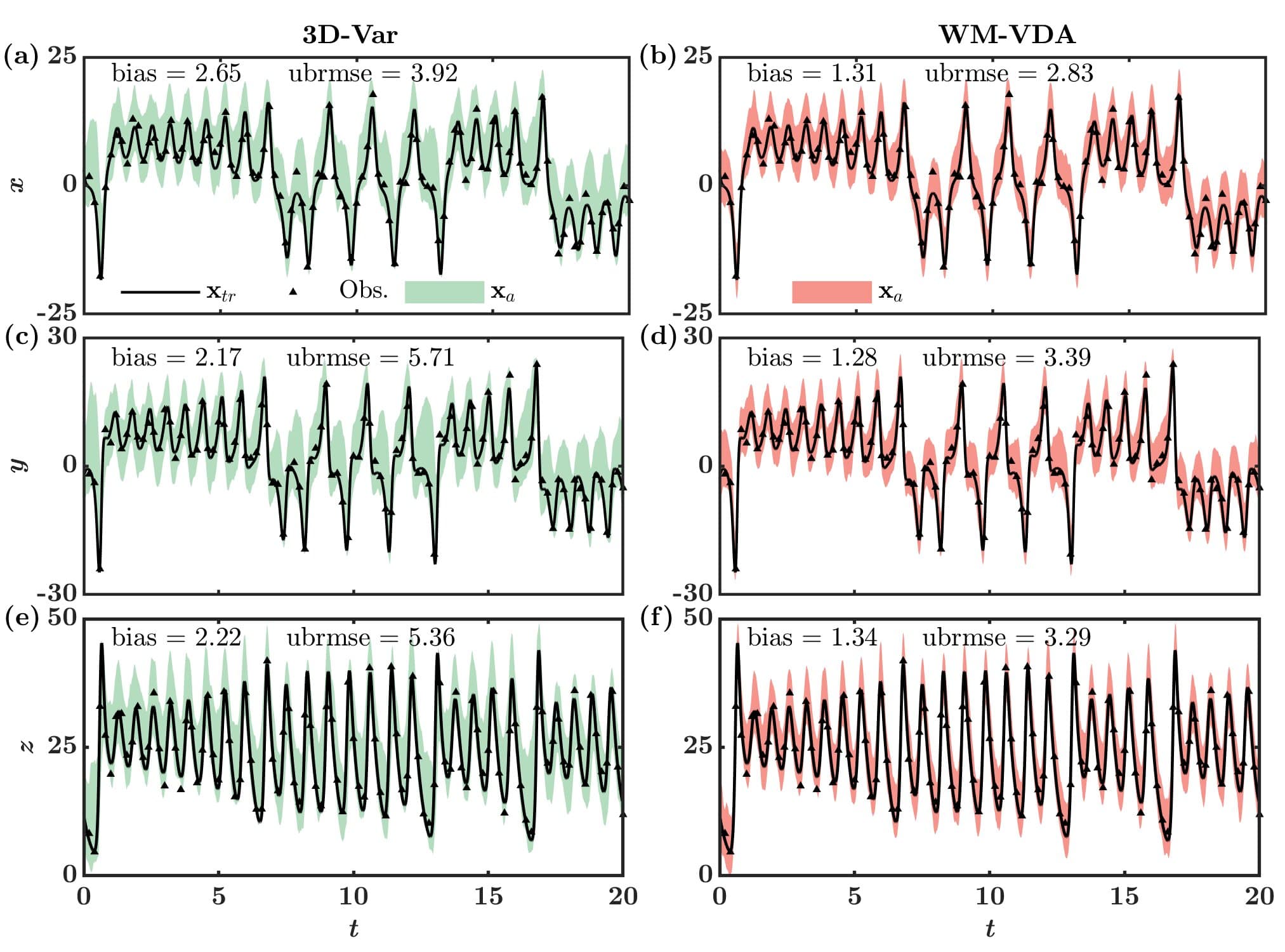}
    \caption{Time evolution of the true state (solid lines) of the Lorenz-63 analysis states, observations (Obs.) for one ensemble member and 2.5 and 97.5 percentiles (shaded areas) of the ensemble members for the 3D-Var (a,\,c,\,e) and the WM-VDA (b,\,d,\,f). The results are obtained for 50 independent ensemble members based on the experimental setup I.}
    \label{fig:5}
\end{figure}

\begin{figure}[h]
    \centering
    \includegraphics[width=1\textwidth]{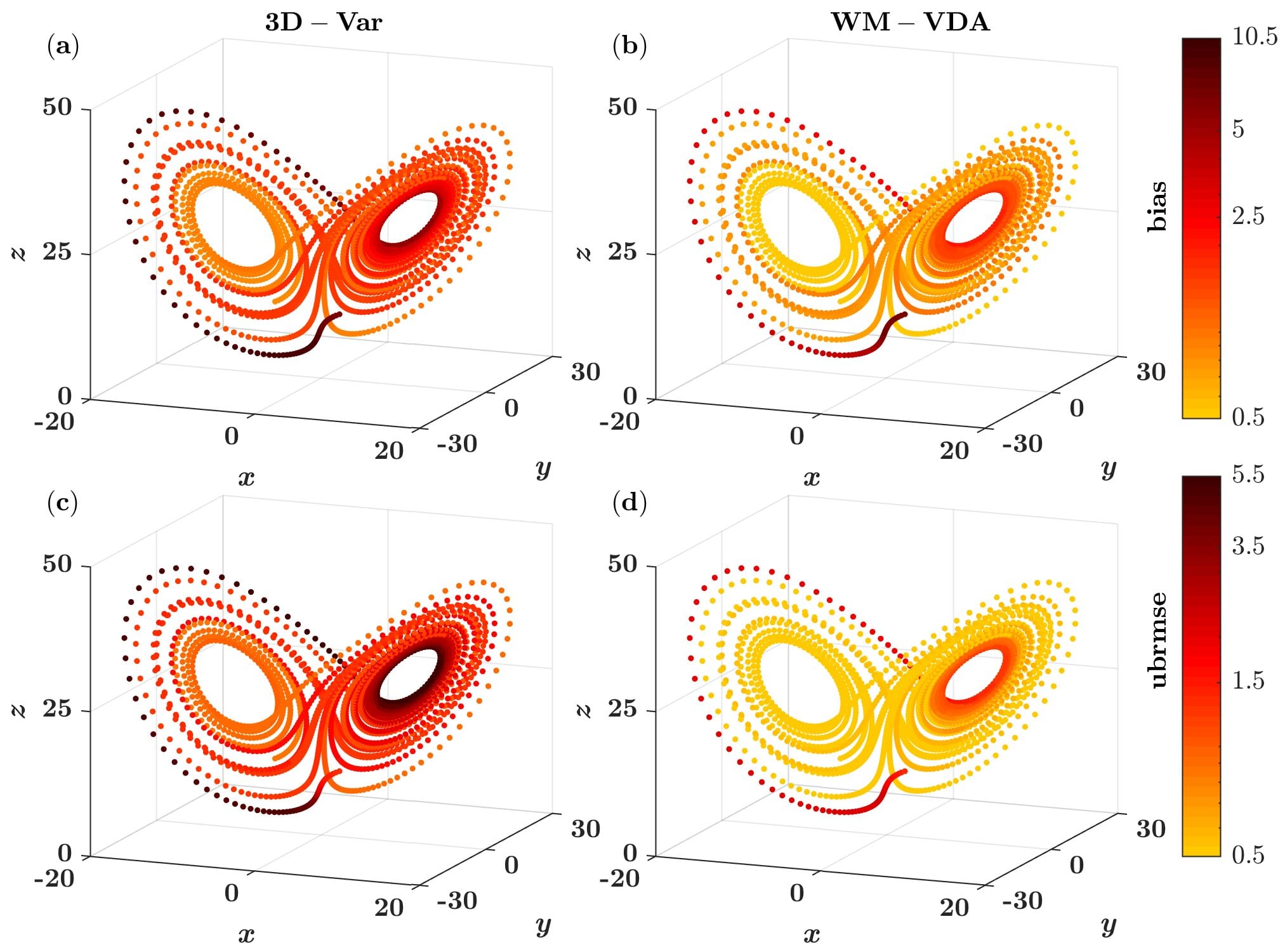}
    \caption{Time evolution of the 95\% confidence bound of the (a,b) bias and (c,d) ubrmse along the phase space of the Lorenz system for the 3D-Var and WM-VDA. The uncertainty range is obtained from 50 independent ensemble members for the first experimental setup.}
    \label{fig:6}
\end{figure}

Figure\,\ref{fig:5} depicts the trajectory of the ground truth and range showing the 2.5 and 97.5 percentiles of the ensemble members by the 3D-Var and WM-VDA schemes for the first experiment. It is seen that for all three state variables, the range for WM-VDA is narrower than the 3D-Var. On average, bias and ubrmse are reduced by 40--50\% and 30--40\%, respectively. Clearly the uncertainty of the quality metrics can also be quantified at each time step for all ensemble members. Under a chaotic dynamics, a narrower range for the error statistics signifies more stable solutions to a biased perturbation of the initial conditions. The time evolution of the uncertainty range, between 2.5 and 97.5 percentiles, of the bias and ubrmse is color coded on the phase space of the Lorenz system in Figure\,\ref{fig:6}. The color gradient from dark red to yellow shows that the uncertainty around the computed error statistics gradually shrinks as the DA progresses and passes the duration of spin-up time. It is seen that the width of uncertainty of the quality metrics is noticeably narrower in the WM-VDA than the 3D-Var. In particular, the range reduces by 47.5\% and 62.1\% for the bias and ubrmse, respectively, demonstrating the advantage of bias-aware WM-VDA over 3D-Var.  Table\,\ref{tab:1} lists the expected values of the bias and ubrmse at the end of both experiments.

\begin{figure}[h]
    \centering
    \includegraphics[width=1\textwidth]{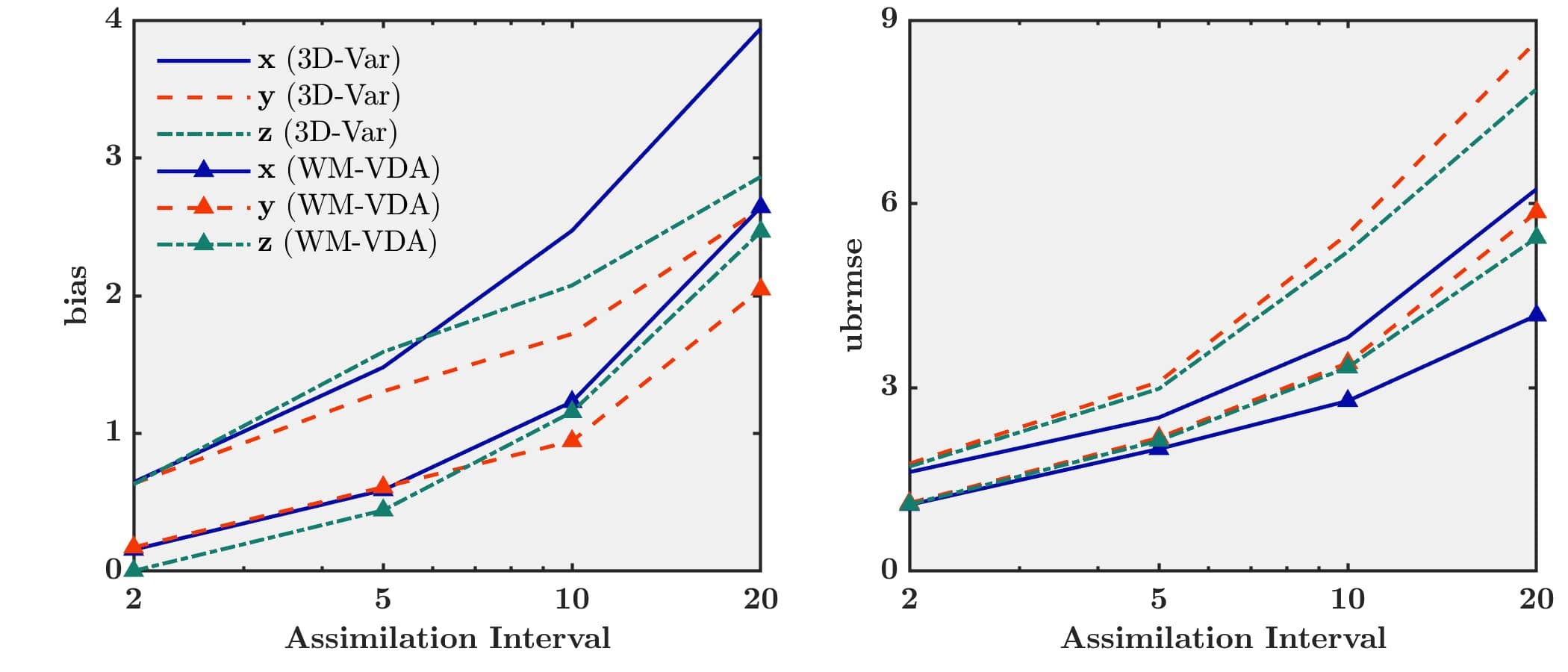}
    \caption{Changes of (a) bias and (b) ubrmse as a function of assimilation interval for 3D-Var and WM-VDA in experimental setup I, where $(x,\,y,\,z)$ represents the three dimensions of the Lorenz system.}
    \label{fig:7}
\end{figure}

To quantify the effects of assimilation intervals on the quality of the DA schemes, experimental setup I was performed for a range of assimilation intervals $T_a=$\{$2\Delta t$,\,$5\Delta t$,\,10$\Delta t$,\,20$\Delta t$\}. The expected values of the obtained quality metrics are shown in Figure\,\ref{fig:7}. As expected, for both schemes, bias and ubrmse increase as the assimilation interval grows and as shown, the WM-VDA outperforms 3D-Var. The gap between the two approaches, in terms of the bias, remains relatively steady even though we see that over the third dimension, it begins to shrink as the assimilation interval increases. However, it appears that as the assimilation interval grows, the gap between the ubrmse keeps increasing. This might have stemmed from a very high value of Lyapunov exponent (0.9) for the Lorenz-63. At such high value, an infinitesimally close trajectories of the state can also deviate significantly as the time progresses. Therefore, for longer assimilation intervals, an analysis state produced by WM-VDA evolves with much less deviation from the true state before the next assimilation cycle and thus exhibit reduced ubrmse than the 3D-Var.

As listed in Table\,\ref{tab:1}, in the second experimental setting, bias in WM-VDA is lower than 3D-Var by 15--50\% whereas there is marginal improvement in the ubrmse. The reason is that the relatively unbiased {\it a priori} probability histogram is selected over a window of time, with a larger uncertainty ($\sigma^2_{x_r}=65-82$) along three dimensions than the first experiment ($\sigma^2_{x_r}=\sqrt{15}$). Thus, its assimilation can only reduce the bias but unable to substantially decrease the ubrmse. However, the results are promising in a sense that even if the spread of {\it a priori} reference probability histogram is much larger than the observations and background probability histogram, the WM-VDA can effectively reduce the bias in analysis without hampering the variance. This observation partially verifies the hypothesis that the WM-VDA provides a new way to effectively assimilate highly uncertain but relatively unbiased probability histogram, obtained from the {\it in situ} data in a climatological sense.

\orange{We also compared the results of setup-II of WM-VDA scheme with the CDF-matching technique implemented on Lorenz-63 system with identical assumptions about error structures as discussed earlier in this section. The CDF-matching scheme is developed without consideration of any {\it a priori} reference information; therefore, in its current form, it is limited to bias correction in either model or observation only. For a fair comparison between the methods, with provision of {\it a priori} information, we deployed the CDF-matching technique by mapping both observations and model state onto the {\it a priori} reference dataset using their respective cumulative histograms. To proceed with CDF-matching, cumulative histogram of the reference dataset is constructed by integrating the model in time from initial state of ($x$, $y$, $z$) = (3,\,$-$3,\,12) with a time step of $\Delta t= 0.01$ [t] over a period of $T = 0-50$ [t] and adding a zero-mean Gaussian noise with covariance $\sqrt3\,\sigma_{b}^2 \mathbf{I}_3$ where, $\sigma_{b}^2 = \sqrt 5$. The observation cumulative histogram is constructed using observations at an interval of 10$\Delta t$ over $T = 0-50$ [t] obtained by corrupting the truth with Gaussian error $v_i \sim \mathcal{N}(\boldsymbol{\beta}_y\mathbb{1}_3,\,\sigma_y^2\, \mathbf{I}_3)$ where $\boldsymbol{\beta}_y=0.15$ and $\sigma_y^2=2$. The model cumulative histogram is computed by propagating the model (Equation \ref{eq:15}) over $T=0-50$ [t] by adding a Gaussian noise $\boldsymbol{\omega}_i \sim \mathcal{N}(\boldsymbol{\beta}_m\mathbb{1}_3,\,\sigma_b^2\, \mathbf{I}_3)$ to the model state at every $\Delta t$, where $\boldsymbol{\beta}_m=0.25$ and $\sigma_b^2= \sqrt 5$. At the end of $T=50$ [t], the model state perturbed with zero-mean Gaussian noise of covariance $\sqrt3\,\sigma_{b}^2 \mathbf{I}_3$ is considered as the initial condition and DA experimentation for CDF-matching technique is then applied upto the next 2000 time steps. At every assimilation time, observation and model are first mapped onto the reference dataset using piece-wise linear CDF-matching to remove bias and then the 3D-Var method was utilized to obtain the analysis state using bias removed model state and observation. It is to note that, in comparison to the WM-VDA, CDF-matching is equipped with more information fed in the form of model and observation cumulative mass functions.}

\orange{It is seen that over 50 independent ensemble simulations, averaged over all three dimensions, CDF-matching reduces the bias by 26\% compared to 28\% by WM-VDA whereas ubrmse increased by 8.6\% compared to overall decrease in ubrmse of 3.3\% by WM-VDA. The difference exists primarily due to the way CDF-matching completely maps the biased source of information onto the relatively unbiased one. Here, since the reference dataset has higher uncertainty in terms of variance, CDF-matching improves bias compared to 3D-Var at the expense of increase in ubrmse. Therefore, if the {\it a priori} information has uncertainty, as can be expected in real condition, WM-VDA can be more effective than a statistic CDF-matching technique in bias correction and also provide a fertile ground to conduct DA experiments in probability domain with reduced uncertainty.}

\section{Summary and Concluding Remarks}

In this study, we discussed the concept of Wasserstein metric (WM) regularization of the variational data assimilation (VDA) techniques, referred to as WM-VDA. In particular, the cost of the classic VDA problem is equipped with a transportation cost that penalizes the mismatch between the probability histograms of the analysis state and a relatively unbiased {\it a priori} reference data, which can be obtained from {\it in-situ} observations over a spatial or temporal window. The presented WM-VDA approach does not need any {\it a priori} knowledge on the sign of the bias and information about its origin from either model and/or observations and automatically retrieve it from the {\it a priori} reference data. We examined the application of WM-VDA for bias removal in a simple first-order linear dynamics and the chaotic Lorenz-63 system \citep{lorenz1963deterministic}. The results demonstrated that WM-VDA can reduce and control propagation of bias under chaotic dynamics. \red{Due to the intrinsic property of the Wasserstein metric to allow natural morphing between the probability histograms \citep{kolouri2017optimal}, a proper value of the regularization parameter alleviates any unrealistic bimodality in the shape of the analysis histogram due to potential biases.} \orange{Initial comparisons with classic CDF matching also demonstrated improved performance, although future research is needed for a comprehensive comparison with other existing methods to fully characterize the advantages and disadvantages of the proposed approach}. 

As it is well understood, in the absence of bias, the classic VDA leads to the lowest possible analysis mean squared error and meets the known Cram\'er-Rao lower bound \citep[CRLB,][]{rao1973linear,cramer1999mathematical}. However, in the presence of systematic biases such a lower bound cannot be met. Even though, we empirically demonstrated that under biased assimilation scenarios, the WM-VDA shows improved performance than the 3D-VAR, future theoretical studies are needed to characterize a closed form expression for such an improvement. \blue{It is to note that, since WM-VDA doesn't need to specifically attribute the bias proportionally to either model or observations, it cannot identify the origin of bias. If such information is needed, future research might be devoted to relate the amount of transported probability masses to the bias of the background state and observations.}

\blue{One of the main challenges of WM-VDA is its high computational complexity. The interior-point optimization algorithm has computational complexity {\it O}($k^3\: \log\:k$) where, $k$ is the number of elements in the support set. This hampers applications of the WM-VDA to large-scale geophysical DA problems. Recent advances in tomographic approximation of the Wasserstein distance \citep{kolouri2018sliced} through slicing the metric via a finite number of 1-D Radon projections can significantly reduce its computational cost for high-dimensions problems and can be a possible direction for future research in geophysical data assimilation.} Moreover, characterization of the regularization parameter through cross validation could be computationally intensive, especially for large-scale Earth system models and new research efforts are required to address this challenge.

As demonstrated empirically, the WM-VDA scheme shows robustness to increased assimilation intervals, which can pave new ways for improving bias-aware satellite VDA systems. Of particular interest is satellite soil moisture or precipitation DA \citep{lin2015dynamical,lin2017combined} as the land and weather models are often biased \citep{reichle2004global,lin2017soil}. \red{To that end, future research is required to understand how WM-VDA can be integrated into land-weather models for assimilation of {\it in-situ} data in probability domain.} \blue {A promising area is to incorporate the Wasserstein metric for bias correction in ensemble-based iterative methods and hybrid variational-ensemble DA techniques. In particular, the theory of optimal mass transport seems to provide a concrete ground to solve the problem of filter degeneracy in particle filters \citep{snyder2008obstacles,van2010nonlinear,reich2015probabilistic,van2019particle}, where the observations and background state do not have overlapping supports and thus the weights cannot be trivially updated using the likelihood function.}


\section*{Acknowledgements}
The first and second authors acknowledge the grant from the National Aeronautics and Space Administration (NASA) Terrestrial Hydrology Program (THP, 80NSSC18K1528) and the New (Early Career) Investigator Program (NIP, 80NSSC18K0742).


\setlength{\tabcolsep}{3pt}
\begin{table}[h]
    \centering
        \caption{Expected values of the bias and ubrmse in 3D-Var and WM-VDA for the experimental setup I and II. Values shown inside of the parentheses are percent reduction of the error metrics due to implementation of the WM-VDA compared to the 3D-Var.}
        \vspace{2mm}
        \begin{tabular}{ccccccc}
        \toprule
        Methods & \multicolumn{3}{c}{bias} & \multicolumn{3}{c}{ ubrmse}\\
        & $x$ & $y$ & $z$ & $x$ & $y$ & $z$\\
        \midrule
        3D-Var & 2.65 & 2.17 & 2.22 & 3.92 & 5.71 & 5.36\\
        setup I: WM-VDA & 1.31 (50.6) & 1.28 (41.0) & 1.34 (39.6) & 2.83 (27.8) & 3.39 (40.6) & 3.29 (38.6)\\
        setup II: WM-VDA & 2.22 (16.2) & 1.67 (23.0) & 1.20 (45.9) & 3.88 (1.0) & 5.42 (5.1) & 5.19 (3.2)\\
        \bottomrule
    \end{tabular}
    \label{tab:1}
\end{table}




\begin{thebibliography}{81}
\providecommand{\natexlab}[1]{#1}
\expandafter\ifx\csname urlstyle\endcsname\relax
  \providecommand{\doi}[1]{doi:\discretionary{}{}{}#1}\else
  \providecommand{\doi}{doi:\discretionary{}{}{}\begingroup
  \urlstyle{rm}\Url}\fi

\bibitem[{\textit{Altman and Gondzio}(1999)}]{altman1999regularized}
Altman, A., and J.~Gondzio (1999), Regularized symmetric indefinite systems in
  interior point methods for linear and quadratic optimization,
  \textit{Optimization Methods and Software}, \textit{11}(1-4), 275--302.

\bibitem[{\textit{Anderson and Anderson}(1999)}]{anderson1999monte}
Anderson, J.~L., and S.~L. Anderson (1999), A monte carlo implementation of the
  nonlinear filtering problem to produce ensemble assimilations and forecasts,
  \textit{Monthly Weather Review}, \textit{127}(12), 2741--2758.

\bibitem[{\textit{Asadi et~al.}(2019)\textit{Asadi, Scott, and
  Clausi}}]{asadi2019data}
Asadi, N., K.~A. Scott, and D.~A. Clausi (2019), Data fusion and data
  assimilation of ice thickness observations using a regularisation framework,
  \textit{Tellus A: Dynamic Meteorology and Oceanography}, pp. 1--20.

\bibitem[{\textit{Asch et~al.}(2016)\textit{Asch, Bocquet, and
  Nodet}}]{asch2016data}
Asch, M., M.~Bocquet, and M.~Nodet (2016), \textit{Data assimilation: methods,
  algorithms, and applications}, vol.~11, SIAM.

\bibitem[{\textit{Aulign{\'e} et~al.}(2007)\textit{Aulign{\'e}, McNally, and
  Dee}}]{auligne2007adaptive}
Aulign{\'e}, T., A.~McNally, and D.~Dee (2007), Adaptive bias correction for
  satellite data in a numerical weather prediction system, \textit{Quarterly
  Journal of the Royal Meteorological Society: A journal of the atmospheric
  sciences, applied meteorology and physical oceanography}, \textit{133}(624),
  631--642.

\bibitem[{\textit{Benamou et~al.}(2015)\textit{Benamou, Carlier, Cuturi, Nenna,
  and Peyr{\'e}}}]{benamou2015iterative}
Benamou, J.-D., G.~Carlier, M.~Cuturi, L.~Nenna, and G.~Peyr{\'e} (2015),
  Iterative bregman projections for regularized transportation problems,
  \textit{SIAM Journal on Scientific Computing}, \textit{37}(2), A1111--A1138.

\bibitem[{\textit{Brenier}(1991)}]{brenier1991polar}
Brenier, Y. (1991), Polar factorization and monotone rearrangement of
  vector-valued functions, \textit{Communications on pure and applied
  mathematics}, \textit{44}(4), 375--417.

\bibitem[{\textit{Charney}(1951)}]{charney1951dynamic}
Charney, J.~G. (1951), Dynamic forecasting by numerical process, in
  \textit{Compendium of meteorology}, pp. 470--482, Springer.

\bibitem[{\textit{Chen et~al.}(2017)\textit{Chen, Georgiou, and
  Tannenbaum}}]{chen2017matrix}
Chen, Y., T.~T. Georgiou, and A.~Tannenbaum (2017), Matrix optimal mass
  transport: a quantum mechanical approach, \textit{IEEE Transactions on
  Automatic Control}, \textit{63}(8), 2612--2619.

\bibitem[{\textit{Chen et~al.}(2018{\natexlab{a}})\textit{Chen, Georgiou, and
  Tannenbaum}}]{chen2018optimal}
Chen, Y., T.~T. Georgiou, and A.~Tannenbaum (2018{\natexlab{a}}), Optimal
  transport for gaussian mixture models, \textit{IEEE Access}, \textit{7},
  6269--6278.

\bibitem[{\textit{Chen et~al.}(2018{\natexlab{b}})\textit{Chen, Georgiou, and
  Tannenbaum}}]{chen2018wasserstein}
Chen, Y., T.~T. Georgiou, and A.~Tannenbaum (2018{\natexlab{b}}), Wasserstein
  geometry of quantum states and optimal transport of matrix-valued measures,
  in \textit{Emerging Applications of Control and Systems Theory}, pp.
  139--150, Springer.

\bibitem[{\textit{Chepurin et~al.}(2005)\textit{Chepurin, Carton, and
  Dee}}]{chepurin2005forecast}
Chepurin, G.~A., J.~A. Carton, and D.~Dee (2005), Forecast model bias
  correction in ocean data assimilation, \textit{Monthly weather review},
  \textit{133}(5), 1328--1342.

\bibitem[{\textit{Courtier et~al.}(1994)\textit{Courtier, Th{\'e}paut, and
  Hollingsworth}}]{courtier1994strategy}
Courtier, P., J.-N. Th{\'e}paut, and A.~Hollingsworth (1994), A strategy for
  operational implementation of 4d-var, using an incremental approach,
  \textit{Quarterly Journal of the Royal Meteorological Society},
  \textit{120}(519), 1367--1387.

\bibitem[{\textit{Cram{\'e}r}(1999)}]{cramer1999mathematical}
Cram{\'e}r, H. (1999), \textit{Mathematical methods of statistics}, vol.~9,
  Princeton university press.

\bibitem[{\textit{Crow et~al.}(2005)\textit{Crow, Koster, Reichle, and
  Sharif}}]{crow2005relevance}
Crow, W.~T., R.~D. Koster, R.~H. Reichle, and H.~O. Sharif (2005), Relevance of
  time-varying and time-invariant retrieval error sources on the utility of
  spaceborne soil moisture products, \textit{Geophysical Research Letters},
  \textit{32}(24).

\bibitem[{\textit{Daley}(1993)}]{daley1993atmospheric}
Daley, R. (1993), \textit{Atmospheric data analysis}, 2, Cambridge university
  press.

\bibitem[{\textit{De~Lannoy et~al.}(2016)\textit{De~Lannoy, de~Rosnay, and
  Reichle}}]{de2016soil}
De~Lannoy, G. J.~M., P.~de~Rosnay, and R.~H. Reichle (2016), Soil moisture data
  assimilation, \textit{Handbook of Hydrometeorological Ensemble Forecasting},
  pp. 1--43.

\bibitem[{\textit{Dee}(2003)}]{dee2003detection}
Dee, D.~P. (2003), Detection and correction of model bias during data
  assimilation, \textit{Meteorological Training Course Lecture Series (ECMWF)}.

\bibitem[{\textit{Dee}(2005)}]{dee2005bias}
Dee, D.~P. (2005), Bias and data assimilation, \textit{Quarterly Journal of the
  Royal Meteorological Society}, \textit{131}(613), 3323--3343.

\bibitem[{\textit{Dee and Da~Silva}(1998)}]{dee1998data}
Dee, D.~P., and A.~M. Da~Silva (1998), Data assimilation in the presence of
  forecast bias, \textit{Quarterly Journal of the Royal Meteorological
  Society}, \textit{124}(545), 269--295.

\bibitem[{\textit{Dee and Uppala}(2009)}]{dee2009variational}
Dee, D.~P., and S.~Uppala (2009), Variational bias correction of satellite
  radiance data in the era-interim reanalysis, \textit{Quarterly Journal of the
  Royal Meteorological Society}, \textit{135}(644), 1830--1841.

\bibitem[{\textit{Dr{\'e}court et~al.}(2006)\textit{Dr{\'e}court, Madsen, and
  Rosbjerg}}]{drecourt2006bias}
Dr{\'e}court, J.-P., H.~Madsen, and D.~Rosbjerg (2006), Bias aware kalman
  filters: Comparison and improvements, \textit{Advances in Water Resources},
  \textit{29}(5), 707--718.

\bibitem[{\textit{Ebtehaj and
  Foufoula-Georgiou}(2013)}]{ebtehaj2013variational}
Ebtehaj, A.~M., and E.~Foufoula-Georgiou (2013), On variational downscaling,
  fusion, and assimilation of hydrometeorological states: A unified framework
  via regularization, \textit{Water Resources Research}, \textit{49}(9),
  5944--5963.

\bibitem[{\textit{Ebtehaj et~al.}(2014)\textit{Ebtehaj, Zupanski, Lerman, and
  Foufoula-Georgiou}}]{ebtehaj2014variational}
Ebtehaj, A.~M., M.~Zupanski, G.~Lerman, and E.~Foufoula-Georgiou (2014),
  Variational data assimilation via sparse regularisation, \textit{Tellus A:
  Dynamic Meteorology and Oceanography}, \textit{66}(1), 21,789.

\bibitem[{\textit{Engquist and Froese}(2013)}]{engquist2013application}
Engquist, B., and B.~D. Froese (2013), Application of the wasserstein metric to
  seismic signals, \textit{arXiv preprint arXiv:1311.4581}.

\bibitem[{\textit{Feyeux et~al.}(2018)\textit{Feyeux, Vidard, and
  Nodet}}]{feyeux2018optimal}
Feyeux, N., A.~Vidard, and M.~Nodet (2018), Optimal transport for variational
  data assimilation, \textit{Nonlinear Processes in Geophysics},
  \textit{25}(1), 55--66.

\bibitem[{\textit{Freitag et~al.}(2010)\textit{Freitag, Nichols, and
  Budd}}]{freitag2010l1}
Freitag, M.~A., N.~K. Nichols, and C.~J. Budd (2010), L1-regularisation for
  ill-posed problems in variational data assimilation, \textit{Pamm},
  \textit{10}(1), 665--668.

\bibitem[{\textit{Gangbo and McCann}(1996)}]{gangbo1996geometry}
Gangbo, W., and R.~J. McCann (1996), The geometry of optimal transportation,
  \textit{Acta Mathematica}, \textit{177}(2), 113--161.

\bibitem[{\textit{Goodliff et~al.}(2015)\textit{Goodliff, Amezcua, and
  Van~Leeuwen}}]{goodliff2015comparing}
Goodliff, M., J.~Amezcua, and P.~J. Van~Leeuwen (2015), Comparing hybrid data
  assimilation methods on the lorenz 1963 model with increasing non-linearity,
  \textit{Tellus A: Dynamic Meteorology and Oceanography}, \textit{67}(1),
  26,928.

\bibitem[{\textit{Hamilton et~al.}(2019)\textit{Hamilton, Berry, and
  Sauer}}]{hamilton2019correcting}
Hamilton, F., T.~Berry, and T.~Sauer (2019), Correcting observation model error
  in data assimilation, \textit{Chaos: An Interdisciplinary Journal of
  Nonlinear Science}, \textit{29}(5), 053,102.

\bibitem[{\textit{Harlim and Hunt}(2007)}]{harlim2007non}
Harlim, J., and B.~R. Hunt (2007), A non-gaussian ensemble filter for
  assimilating infrequent noisy observations, \textit{Tellus A: Dynamic
  Meteorology and Oceanography}, \textit{59}(2), 225--237.

\bibitem[{\textit{Hazan et~al.}(2017)\textit{Hazan, Singh, and
  Zhang}}]{hazan2017learning}
Hazan, E., K.~Singh, and C.~Zhang (2017), Learning linear dynamical systems via
  spectral filtering, in \textit{Advances in Neural Information Processing
  Systems}, pp. 6702--6712.

\bibitem[{\textit{Hazan et~al.}(2018)\textit{Hazan, Lee, Singh, Zhang, and
  Zhang}}]{hazan2018spectral}
Hazan, E., H.~Lee, K.~Singh, C.~Zhang, and Y.~Zhang (2018), Spectral filtering
  for general linear dynamical systems, in \textit{Advances in Neural
  Information Processing Systems}, pp. 4634--4643.

\bibitem[{\textit{Jin et~al.}(2019)\textit{Jin, Lin, Segers, Xie, and
  Heemink}}]{jin2019machine}
Jin, J., H.~X. Lin, A.~Segers, Y.~Xie, and A.~Heemink (2019), Machine learning
  for observation bias correction with application to dust storm data
  assimilation, \textit{Atmospheric Chemistry and Physics}, \textit{19}(15),
  10,009--10,026.

\bibitem[{\textit{Kalman}(1960)}]{kalman1960new}
Kalman, R.~E. (1960), A new approach to linear filtering and prediction
  problems, \textit{Journal of basic Engineering}, \textit{82}(1), 35--45.

\bibitem[{\textit{Kalnay}(2003)}]{kalnay2003atmospheric}
Kalnay, E. (2003), \textit{Atmospheric modeling, data assimilation and
  predictability}, Cambridge university press.

\bibitem[{\textit{Kalnay et~al.}(2007)\textit{Kalnay, Li, Miyoshi, Yang, and
  Ballabrera-Poy}}]{kalnay20074}
Kalnay, E., H.~Li, T.~Miyoshi, S.-C. Yang, and J.~Ballabrera-Poy (2007),
  4-d-var or ensemble kalman filter?, \textit{Tellus A: Dynamic Meteorology and
  Oceanography}, \textit{59}(5), 758--773.

\bibitem[{\textit{Kantorovich}(1942)}]{kantorovich1942translocation}
Kantorovich, L.~V. (1942), On the translocation of masses, in \textit{Dokl.
  Akad. Nauk. USSR (NS)}, vol.~37, pp. 199--201.

\bibitem[{\textit{Kolouri et~al.}(2017)\textit{Kolouri, Park, Thorpe, Slepcev,
  and Rohde}}]{kolouri2017optimal}
Kolouri, S., S.~R. Park, M.~Thorpe, D.~Slepcev, and G.~K. Rohde (2017), Optimal
  mass transport: Signal processing and machine-learning applications,
  \textit{IEEE signal processing magazine}, \textit{34}(4), 43--59.

\bibitem[{\textit{Kolouri et~al.}(2018)\textit{Kolouri, Rohde, and
  Hoffmann}}]{kolouri2018sliced}
Kolouri, S., G.~K. Rohde, and H.~Hoffmann (2018), Sliced wasserstein distance
  for learning gaussian mixture models, in \textit{Proceedings of the IEEE
  Conference on Computer Vision and Pattern Recognition}, pp. 3427--3436.

\bibitem[{\textit{Kumar et~al.}(2009)\textit{Kumar, Reichle, Koster, Crow, and
  Peters-Lidard}}]{kumar2009role}
Kumar, S.~V., R.~H. Reichle, R.~D. Koster, W.~T. Crow, and C.~D. Peters-Lidard
  (2009), Role of subsurface physics in the assimilation of surface soil
  moisture observations, \textit{Journal of hydrometeorology}, \textit{10}(6),
  1534--1547.

\bibitem[{\textit{Kumar et~al.}(2012)\textit{Kumar, Reichle, Harrison,
  Peters-Lidard, Yatheendradas, and Santanello}}]{kumar2012comparison}
Kumar, S.~V., R.~H. Reichle, K.~W. Harrison, C.~D. Peters-Lidard,
  S.~Yatheendradas, and J.~A. Santanello (2012), A comparison of methods for a
  priori bias correction in soil moisture data assimilation, \textit{Water
  Resources Research}, \textit{48}(3).

\bibitem[{\textit{Kutta}(1901)}]{kutta1901beitrag}
Kutta, W. (1901), Beitrag zur naherungsweisen integration totaler
  differentialgleichungen, \textit{Z. Math. Phys.}, \textit{46}, 435--453.

\bibitem[{\textit{Law and Stuart}(2012)}]{law2012evaluating}
Law, K.~J., and A.~M. Stuart (2012), Evaluating data assimilation algorithms,
  \textit{Monthly Weather Review}, \textit{140}(11), 3757--3782.

\bibitem[{\textit{Leith}(1993)}]{leith1993numerical}
Leith, C. (1993), Numerical models of weather and climate, \textit{Plasma
  physics and controlled fusion}, \textit{35}(8), 919.

\bibitem[{\textit{Li et~al.}(2013)\textit{Li, Wang, and Zhang}}]{li2013novel}
Li, P., Q.~Wang, and L.~Zhang (2013), A novel earth mover's distance
  methodology for image matching with gaussian mixture models, in
  \textit{Proceedings of the IEEE International Conference on Computer Vision},
  pp. 1689--1696.

\bibitem[{\textit{Lin et~al.}(2015)\textit{Lin, Ebtehaj, Bras, Flores, and
  Wang}}]{lin2015dynamical}
Lin, L.-F., A.~M. Ebtehaj, R.~L. Bras, A.~N. Flores, and J.~Wang (2015),
  Dynamical precipitation downscaling for hydrologic applications using wrf
  4d-var data assimilation: Implications for gpm era, \textit{Journal of
  Hydrometeorology}, \textit{16}(2), 811--829.

\bibitem[{\textit{Lin et~al.}(2017{\natexlab{a}})\textit{Lin, Ebtehaj, Flores,
  Bastola, and Bras}}]{lin2017combined}
Lin, L.-F., A.~M. Ebtehaj, A.~N. Flores, S.~Bastola, and R.~L. Bras
  (2017{\natexlab{a}}), Combined assimilation of satellite precipitation and
  soil moisture: A case study using trmm and smos data, \textit{Monthly Weather
  Review}, \textit{145}(12), 4997--5014.

\bibitem[{\textit{Lin et~al.}(2017{\natexlab{b}})\textit{Lin, Ebtehaj, Wang,
  and Bras}}]{lin2017soil}
Lin, L.-F., A.~M. Ebtehaj, J.~Wang, and R.~L. Bras (2017{\natexlab{b}}), Soil
  moisture background error covariance and data assimilation in a coupled
  land-atmosphere model, \textit{Water Resources Research}, \textit{53}(2),
  1309--1335.

\bibitem[{\textit{Liu et~al.}(2018)\textit{Liu, Reichle, Bindlish, Cosh, Crow,
  de~Jeu, De~Lannoy, Huffman, and Jackson}}]{liu2018contributions}
Liu, Q., R.~H. Reichle, R.~Bindlish, M.~H. Cosh, W.~T. Crow, R.~de~Jeu, G.~J.
  De~Lannoy, G.~J. Huffman, and T.~J. Jackson (2018), The contributions of
  precipitation and soil moisture observations to the skill of soil moisture
  estimates in a land data assimilation system, \textit{Journal of
  Hydrometeorology}, \textit{19}(2).

\bibitem[{\textit{Lorenc}(1986)}]{lorenc1986analysis}
Lorenc, A.~C. (1986), Analysis methods for numerical weather prediction,
  \textit{Quarterly Journal of the Royal Meteorological Society},
  \textit{112}(474), 1177--1194.

\bibitem[{\textit{Lorenc et~al.}(2015)\textit{Lorenc, Bowler, Clayton, Pring,
  and Fairbairn}}]{lorenc2015comparison}
Lorenc, A.~C., N.~E. Bowler, A.~M. Clayton, S.~R. Pring, and D.~Fairbairn
  (2015), Comparison of hybrid-4denvar and hybrid-4dvar data assimilation
  methods for global nwp, \textit{Monthly Weather Review}, \textit{143}(1),
  212--229.

\bibitem[{\textit{Lorenz}(1963)}]{lorenz1963deterministic}
Lorenz, E.~N. (1963), Deterministic nonperiodic flow, \textit{Journal of the
  atmospheric sciences}, \textit{20}(2), 130--141.

\bibitem[{\textit{Miller et~al.}(1994)\textit{Miller, Ghil, and
  Gauthiez}}]{miller1994advanced}
Miller, R.~N., M.~Ghil, and F.~Gauthiez (1994), Advanced data assimilation in
  strongly nonlinear dynamical systems, \textit{Journal of the atmospheric
  sciences}, \textit{51}(8), 1037--1056.

\bibitem[{\textit{Monge}(1781)}]{monge1781memoire}
Monge, G. (1781), M{\'e}moire sur la th{\'e}orie des d{\'e}blais et des
  remblais, \textit{Histoire de l'Acad{\'e}mie Royale des Sciences de Paris}.

\bibitem[{\textit{Motamed and Appelo}(2018)}]{motamed2018wasserstein}
Motamed, M., and D.~Appelo (2018), Wasserstein metric-driven bayesian inversion
  with application to wave propagation problems, \textit{arXiv preprint
  arXiv:1807.09682}.

\bibitem[{\textit{Ning et~al.}(2014)\textit{Ning, Carli, Ebtehaj,
  Foufoula-Georgiou, and Georgiou}}]{ning2014coping}
Ning, L., F.~P. Carli, A.~M. Ebtehaj, E.~Foufoula-Georgiou, and T.~T. Georgiou
  (2014), Coping with model error in variational data assimilation using
  optimal mass transport, \textit{Water Resources Research}, \textit{50}(7),
  5817--5830.

\bibitem[{\textit{Pauwels et~al.}(2013)\textit{Pauwels, De~Lannoy,
  Hendricks~Franssen, and Vereecken}}]{pauwels2013simultaneous}
Pauwels, V.~R., G.~J. De~Lannoy, H.-J. Hendricks~Franssen, and H.~Vereecken
  (2013), Simultaneous estimation of model state variables and observation and
  forecast biases using a two-stage hybrid kalman filter, \textit{Hydrology and
  Earth System Sciences}, \textit{17}(9), 3499--3521.

\bibitem[{\textit{Peyr{\'e} et~al.}(2019)\textit{Peyr{\'e}, Cuturi
  et~al.}}]{peyre2019computational}
Peyr{\'e}, G., M.~Cuturi, et~al. (2019), Computational optimal transport,
  \textit{Foundations and Trends{\textregistered} in Machine Learning},
  \textit{11}(5-6), 355--607.

\bibitem[{\textit{Rabier}(2005)}]{rabier2005overview}
Rabier, F. (2005), Overview of global data assimilation developments in
  numerical weather-prediction centres, \textit{Quarterly Journal of the Royal
  Meteorological Society}, \textit{131}(613), 3215--3233.

\bibitem[{\textit{Radakovich et~al.}(2001)\textit{Radakovich, Houser, da~Silva,
  and Bosilovich}}]{radakovich2001results}
Radakovich, J.~D., P.~R. Houser, A.~da~Silva, and M.~G. Bosilovich (2001),
  Results from global land-surface data assimilation methods, in \textit{AGU
  Spring Meeting Abstracts}.

\bibitem[{\textit{Rao et~al.}(1973)\textit{Rao, Rao, Statistiker, Rao, and
  Rao}}]{rao1973linear}
Rao, C.~R., C.~R. Rao, M.~Statistiker, C.~R. Rao, and C.~R. Rao (1973),
  \textit{Linear statistical inference and its applications}, vol.~2, Wiley New
  York.

\bibitem[{\textit{Reich}(2012)}]{reich2012gaussian}
Reich, S. (2012), A gaussian-mixture ensemble transform filter,
  \textit{Quarterly Journal of the Royal Meteorological Society},
  \textit{138}(662), 222--233.

\bibitem[{\textit{Reich and Cotter}(2015)}]{reich2015probabilistic}
Reich, S., and C.~Cotter (2015), \textit{Probabilistic forecasting and Bayesian
  data assimilation}, Cambridge University Press.

\bibitem[{\textit{Reichle and Koster}(2004)}]{reichle2004bias}
Reichle, R.~H., and R.~D. Koster (2004), Bias reduction in short records of
  satellite soil moisture, \textit{Geophysical Research Letters},
  \textit{31}(19).

\bibitem[{\textit{Reichle et~al.}(2004)\textit{Reichle, Koster, Dong, and
  Berg}}]{reichle2004global}
Reichle, R.~H., R.~D. Koster, J.~Dong, and A.~A. Berg (2004), Global soil
  moisture from satellite observations, land surface models, and ground data:
  Implications for data assimilation, \textit{Journal of Hydrometeorology},
  \textit{5}(3), 430--442.

\bibitem[{\textit{Reichle et~al.}(2007)\textit{Reichle, Koster, Liu, Mahanama,
  Njoku, and Owe}}]{reichle2007comparison}
Reichle, R.~H., R.~D. Koster, P.~Liu, S.~P. Mahanama, E.~G. Njoku, and M.~Owe
  (2007), Comparison and assimilation of global soil moisture retrievals from
  the advanced microwave scanning radiometer for the earth observing system
  (amsr-e) and the scanning multichannel microwave radiometer (smmr),
  \textit{Journal of Geophysical Research: Atmospheres}, \textit{112}(D9).

\bibitem[{\textit{Reichle et~al.}(2010)\textit{Reichle, Kumar, Mahanama,
  Koster, and Liu}}]{reichle2010assimilation}
Reichle, R.~H., S.~V. Kumar, S.~P. Mahanama, R.~D. Koster, and Q.~Liu (2010),
  Assimilation of satellite-derived skin temperature observations into land
  surface models, \textit{Journal of Hydrometeorology}, \textit{11}(5),
  1103--1122.

\bibitem[{\textit{Rubner et~al.}(2000)\textit{Rubner, Tomasi, and
  Guibas}}]{rubner2000earth}
Rubner, Y., C.~Tomasi, and L.~J. Guibas (2000), The earth mover's distance as a
  metric for image retrieval, \textit{International journal of computer
  vision}, \textit{40}(2), 99--121.

\bibitem[{\textit{Runge}(1895)}]{runge1895numerische}
Runge, C. (1895), {\"U}ber die numerische aufl{\"o}sung von
  differentialgleichungen, \textit{Mathematische Annalen}, \textit{46}(2),
  167--178.

\bibitem[{\textit{Santambrogio}(2015)}]{santambrogio2015optimal}
Santambrogio, F. (2015), Optimal transport for applied mathematicians,
  \textit{Birk{\"a}user, NY}, \textit{55}, 58--63.

\bibitem[{\textit{Snyder et~al.}(2008)\textit{Snyder, Bengtsson, Bickel, and
  Anderson}}]{snyder2008obstacles}
Snyder, C., T.~Bengtsson, P.~Bickel, and J.~Anderson (2008), Obstacles to
  high-dimensional particle filtering, \textit{Monthly Weather Review},
  \textit{136}(12), 4629--4640.

\bibitem[{\textit{Van~Leeuwen}(2010)}]{van2010nonlinear}
Van~Leeuwen, P.~J. (2010), Nonlinear data assimilation in geosciences: an
  extremely efficient particle filter, \textit{Quarterly Journal of the Royal
  Meteorological Society}, \textit{136}(653), 1991--1999.

\bibitem[{\textit{Van~Leeuwen et~al.}(2019)\textit{Van~Leeuwen, K{\"u}nsch,
  Nerger, Potthast, and Reich}}]{van2019particle}
Van~Leeuwen, P.~J., H.~R. K{\"u}nsch, L.~Nerger, R.~Potthast, and S.~Reich
  (2019), Particle filters for high-dimensional geoscience applications: A
  review, \textit{Quarterly Journal of the Royal Meteorological Society}.

\bibitem[{\textit{Villani}(2003)}]{villani2003topics}
Villani, C. (2003), \textit{Topics in optimal transportation}, 58, American
  Mathematical Soc.

\bibitem[{\textit{Villani}(2008)}]{villani2008optimal}
Villani, C. (2008), \textit{Optimal transport: old and new}, vol. 338, Springer
  Science \& Business Media.

\bibitem[{\textit{Villarini et~al.}(2008)\textit{Villarini, Mandapaka,
  Krajewski, and Moore}}]{villarini2008rainfall}
Villarini, G., P.~V. Mandapaka, W.~F. Krajewski, and R.~J. Moore (2008),
  Rainfall and sampling uncertainties: A rain gauge perspective,
  \textit{Journal of Geophysical Research: Atmospheres}, \textit{113}(D11).

\bibitem[{\textit{Wahba and Wendelberger}(1980)}]{wahba1980some}
Wahba, G., and J.~Wendelberger (1980), Some new mathematical methods for
  variational objective analysis using splines and cross validation,
  \textit{Monthly weather review}, \textit{108}(8), 1122--1143.

\bibitem[{\textit{Wang et~al.}(2008)\textit{Wang, Barker, Snyder, and
  Hamill}}]{wang2008hybrid}
Wang, X., D.~M. Barker, C.~Snyder, and T.~M. Hamill (2008), A hybrid
  etkf--3dvar data assimilation scheme for the wrf model. part i: Observing
  system simulation experiment, \textit{Monthly Weather Review},
  \textit{136}(12), 5116--5131.

\bibitem[{\textit{Zhu et~al.}(2014)\textit{Zhu, Derber, Collard, Dee, Treadon,
  Gayno, and Jung}}]{zhu2014enhanced}
Zhu, Y., J.~Derber, A.~Collard, D.~Dee, R.~Treadon, G.~Gayno, and J.~A. Jung
  (2014), Enhanced radiance bias correction in the national centers for
  environmental prediction's gridpoint statistical interpolation data
  assimilation system, \textit{Quarterly Journal of the Royal Meteorological
  Society}, \textit{140}(682), 1479--1492.

\bibitem[{\textit{Zupanski}(1997)}]{zupanski1997general}
Zupanski, D. (1997), A general weak constraint applicable to operational 4dvar
  data assimilation systems, \textit{Monthly Weather Review}, \textit{125}(9),
  2274--2292.

\end{thebibliography}
\end{document}